\documentclass[a4paper]{article}
\pdfoutput=1
\usepackage[a4paper]{geometry}
\usepackage{amsmath}
\usepackage{color}
\usepackage{graphicx}
\usepackage{caption}
\usepackage{subcaption}
\usepackage{tabularx}
\usepackage{booktabs}
\usepackage{multirow}
\usepackage{authblk}
\usepackage{cite}
\usepackage[utf8]{inputenc}
\usepackage[hidelinks]{hyperref}

\graphicspath{{figs/}}

\title{\bf Discovering and profiling \texorpdfstring{$Z'$}{Z'} bosons using asymmetry observables in top quark pair production with the lepton-plus-jets final state at the LHC}

\author[1]{Lucio Cerrito\thanks{E-mail: {\tt lucio.cerrito@cern.ch}}}
\author[2,3]{Declan Millar\thanks{E-mail: {\tt declan.millar@cern.ch}}}
\author[3]{Stefano Moretti\thanks{E-mail: {\tt S.Moretti@soton.ac.uk}}}
\author[4]{Francesco Span\`{o}\thanks{E-mail: {\tt francesco.spano@cern.ch}}}

\affil[1]{\it Department of Physics, University of Rome Tor Vergata and INFN, Via Della Ricerca Scientifica, 1, Rome, 00133, Italy}
\affil[2]{\it School of Physics and Astronomy, Queen Mary University of London, Mile End Road, London E1 4NS, United Kingdom}
\affil[3]{\it School of Physics and Astronomy, University of Southampton, Highfield, Southampton SO17 1BJ, United Kingdom}
\affil[4]{\it Department of Physics, Royal Holloway University of London, Egham Hill, Egham TW20 0EX, United Kingdom}

\date{}

\begin{document}

\pagenumbering{Alph}

\begin{titlepage}

  \maketitle

  \begin{abstract}
    The sensitivity of top quark pair production at the Large Hadron Collider to the presence and nature of an underlying $Z'$ boson is studied, accounting for six-fermion decay with full tree-level Standard Model $t\bar{t}$ interference and all intermediate particles allowed off-shell. Focus is placed on the lepton-plus-jets final state, emulating experimental conditions, including kinematic requirements and top quark pair reconstruction in the presence of missing transverse energy and combinatorial ambiguity in jet-top assignment. Considering a resonance with mass of $4$~TeV, and assuming $300$~fb$^{-1}$ of proton-proton collisions with a centre of mass energy of $13$~TeV data, a combination of forward-backward and top polarisation asymmetries are shown to distinguish $Z'$ embedded by different classes of Grand-Unified-Theory-inspired models. In combination with the differential cross section, they may be used to increase the significance of the signal when tested against the Standard Model, as shown using a likelihood-based statistical test.
  \end{abstract}

  \thispagestyle{empty}

\end{titlepage}

\pagenumbering{arabic}

\section{Introduction}
\label{sec:introduction}

New fundamental, massive, neutral, spin-1 gauge bosons ($Z'$) appear ubiquitously in theories that extend gauge and/or spacetime symmetries Beyond the Standard Model (BSM). These $Z'$ typically arise due to residual U$(1)$ gauge symmetries after the spontaneous symmetry breaking of a Grand Unified Theory (GUT).\footnotemark{} As energies lower this leads to a simple U$(1)$ gauge extension of the Standard Model (SM) symmetry group~\cite{dib1987}, which is then spontaneously broken to return the SM. Furthermore, $Z'$ may also arise in extra-dimensional models as Kaluza-Klein excitations of SM gauge fields, leading to new, quasi-degenerate resonances in collider experiments~\cite{arkani-hamed1998,antoniadis1998,randall1999,accomando2013}.

\footnotetext{With the notable exception of SU$(5)$, which has the same rank as G$_{SM}$ (Fig.~\ref{fig:symmetry_breaking_chart}).}

Experimentally, a $Z'$ would label any additional neutral resonance appearing in, for example, a lepton-antilepton pair ($\ell^{+}\ell^{-}$) or top-antitop quark pair ($t\bar{t}$) mass spectrum. The strongest limits for such a state exist for the former signature, known as Drell-Yan (DY), with $l=e,\mu$.\footnotemark{} The experimental signature for DY is clean and theoretical uncertainties for inclusive quantities are small, including those associated to higher order effects, both two-loop from Quantum Chromo-Dynamics (QCD) and one-loop Electro-Weak (EW) ones~\cite{altarelli1979, hamberg1991, harlander2002, anastasiou2004, alekhin2006, dittmaier2010}. Therefore, theories featuring generationally universal couplings to the new $Z'$ are best discovered using dilepton observables, such as the cross section profiled in the dilepton invariant mass. Charge asymmetries can additionally be used to measure $Z'$ properties and couplings to SM particles in the DY channel, with potential as a discovery tool in certain scenarios~\cite{accomando2016}.

\footnotetext{Present limits are discussed in Sec.~\ref{sec:summary_and_present_limits}.}

\begin{figure}
  \centering
  \begin{subfigure}{0.325\textwidth}
    \centering
    \includegraphics[width=0.9\textwidth]{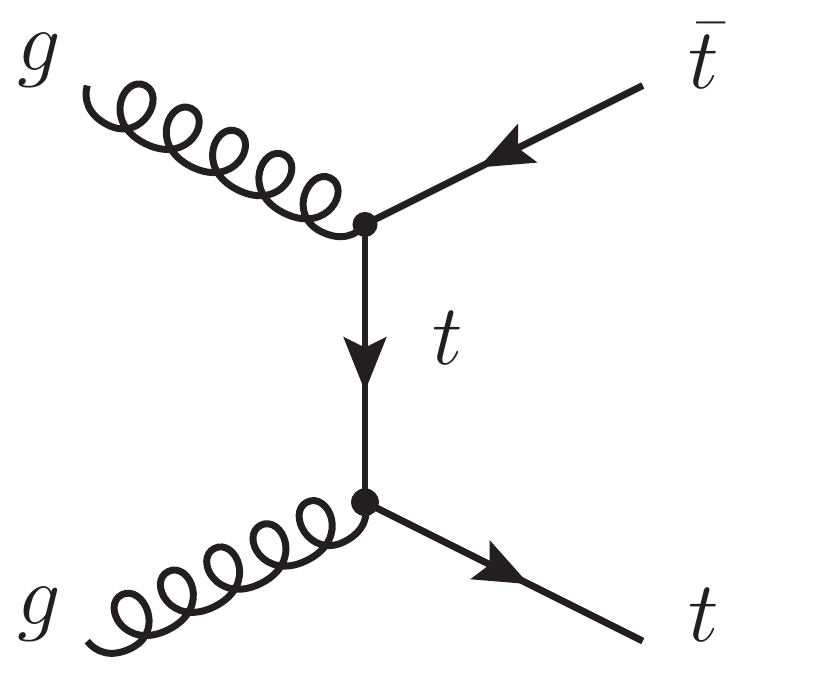}
    \caption*{}
    \vspace{5mm}
  \end{subfigure}
  \begin{subfigure}{0.325\textwidth}
    \centering
    \includegraphics[width=0.9\textwidth]{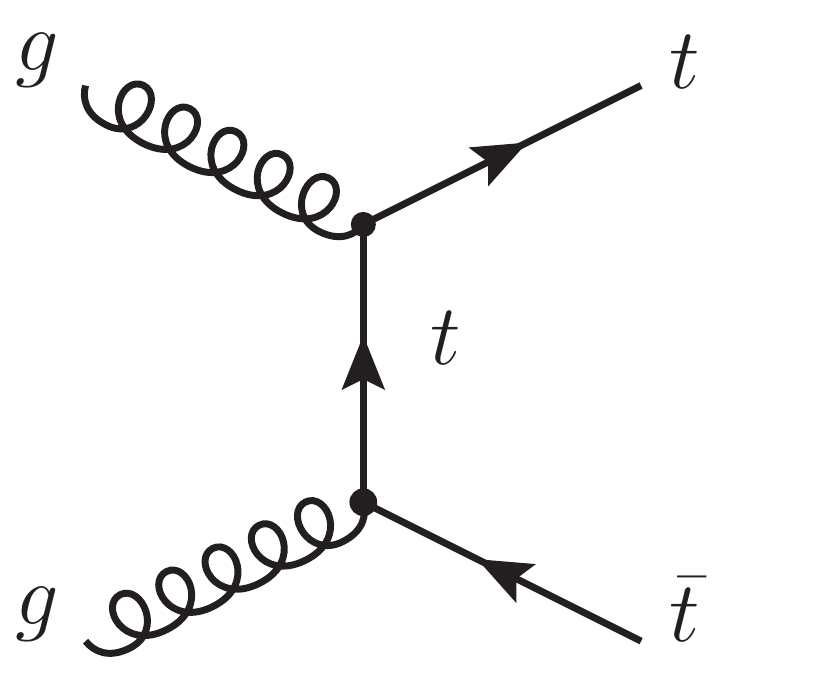}
    \caption{}
    \vspace{5mm}
  \end{subfigure}
  \begin{subfigure}{0.325\textwidth}
    \centering
    \includegraphics[width=0.9\textwidth]{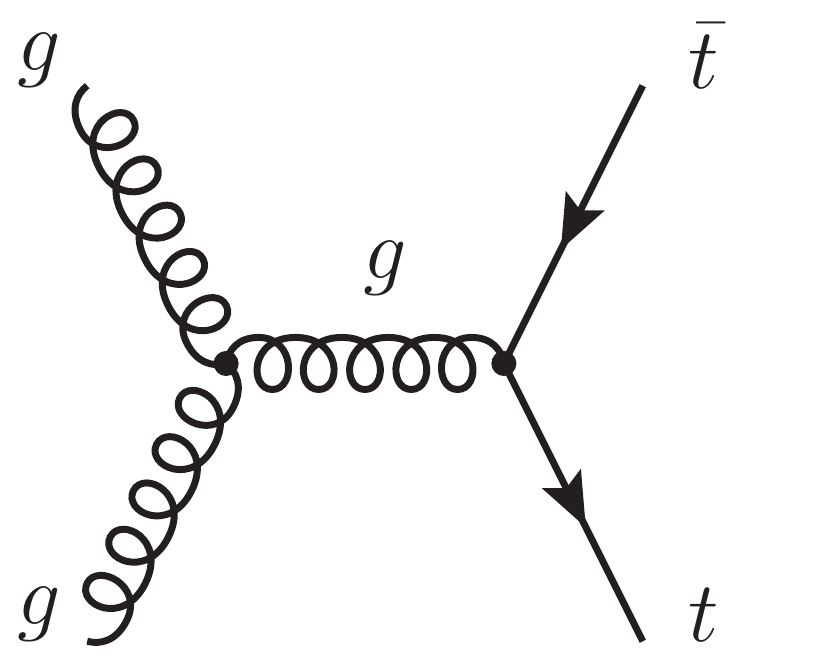}
    \caption*{}
    \vspace{5mm}
  \end{subfigure}
  \begin{subfigure}{0.325\textwidth}
    \centering
    \includegraphics[width=0.9\textwidth]{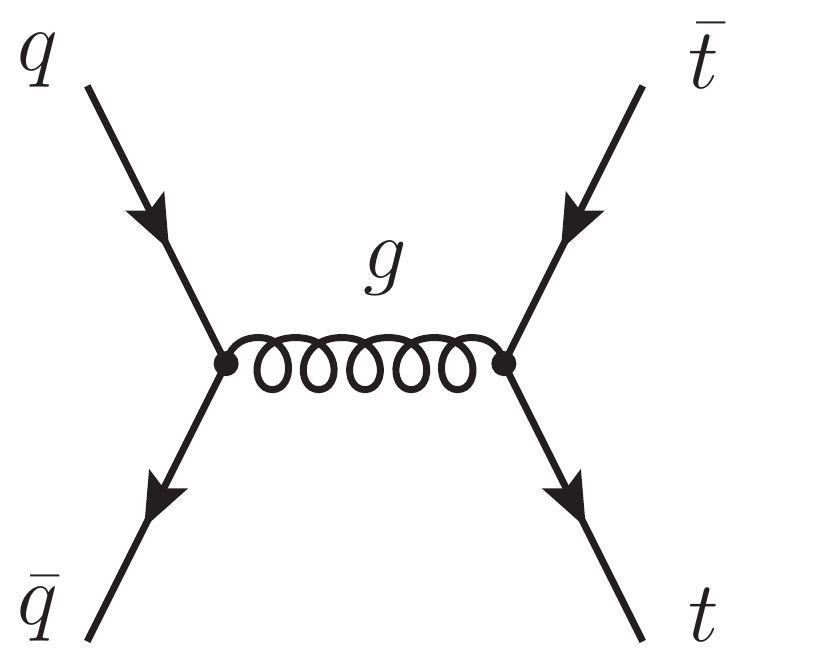}
    \caption{}
  \end{subfigure}
  \begin{subfigure}{0.325\textwidth}
    \centering
    \includegraphics[width=0.9\textwidth]{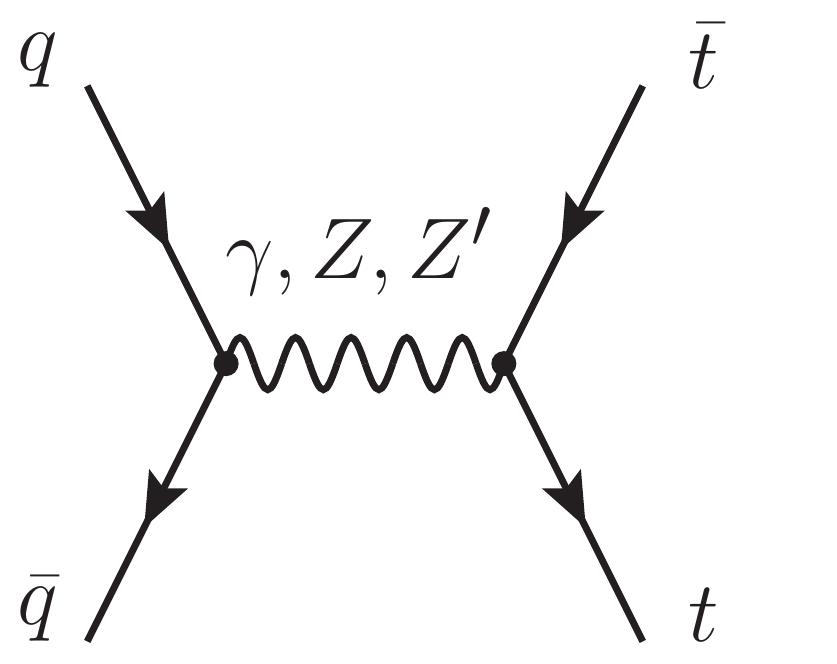}
    \caption{}
  \end{subfigure}
  \caption{Leading order Feynman diagrams for top pair production in BSM:
  (a) gluon-gluon fusion;
  (b) quark-antiquark annihilation to a gluon;
  (c) quark-antiquark annihilation to a photon, $Z$ boson or $Z'$ boson.}
  \label{fig:lottdiagrams}
\end{figure}

The top-antitop pair is an alternative observation channel for $Z'$ bosons~\cite{beneke2000, han2008, bernreuther2008, frederix2009}. Its reduced importance for the discovery of a $Z'$ is due to the larger background, which includes irreducible, QCD production of $t\bar{t}$ pairs (see Fig.~\ref{fig:lottdiagrams}), in addition to the EW irreducible background, for each top decay signature. Furthermore, there are reducible backgrounds, especially severe in the fully hadronic decay mode. The complex 6-body final state results in either multi-jet signatures or events with one or multiple sources of missing transverse energy (because of neutrinos escaping detection). This reduces the potential for first discovery in $t\bar t$ compared to the DY channel. However, there are theoretical scenarios that favour strong couplings between a new $Z'$ and top quarks. These can arise due to leptophobic $Z'$, or extra bosons with an enhanced coupling to third generation fermions, as is common in, for example, Composite Higgs Models~\cite{barducci2012}.

In addition to their importance in extracting the couplings to top quarks, resonance searches in the $t\bar{t}$ channel can offer additional handles on the properties of a $Z'$. This is due to unique variables available for this channel, owing to the fact that (anti)top quarks decay prior to hadronisation, meaning spin information is effectively transmitted to their decay products. The couplings of initial and final state fermions to the $Z'$ are different between charge/spin asymmetries and cross sections. This implies that charge/spin asymmetries provide additional information on the nature of the underlying $Z'$. However, defining these asymmetries requires the reconstruction of the top-antitop quark pair. It is difficult to fully reconstruct the system in dilepton $t\bar{t}$ events which feature two sources of missing transverse energy, while fully hadronic $t\bar{t}$ events come with very large backgrounds. For these reasons, the present study is limited to semileptonic top pair decays, leading to the lepton-plus-jets final state, which permits well reconstructed asymmetries while avoiding overwhelming background processes.

In this paper the same range of models studied in Ref.~\cite{basso2012a} is explored. That analysis is extended here to include the off-shell semileptonic decay of the top quarks and emulate the consequently required reconstruction and corresponding decay-level observables. Numerous phenomenological papers have been produced that study the same or similar observables at the Large Hadron Collider (LHC) for a range of different BSM scenarios~\cite{godbole2010c, godbole2010b, fajfer2012, godbole2015}, including those accounting for full showering/hadronisation and fast simulation of detector effects~\cite{hubaut2005, berger2011, baumgart2013}. Other studies focus on different observables, particularly those suited to fully hadronic or dileptonic top decays~\cite{baumgart2011,tweedie2014}. In addition to the specific selection of GUT derived models explored in this study, the observables described in Sec.~\ref{sec:method} are uniquely used in combination with the differential cross section, in a two dimensional analysis, and the subsequent improvement in the statistical significance is calculated. It is shown how a combination of these observables enables different classes of models to be distinguished between.

The paper is organised as follows: the sensitivity to the presence of a single $Z'$ boson at the LHC arising from a number of generationally universal benchmark models is explored, with these Models described in Sec.~\ref{sec:models}. Next the method employed in this analysis is outlined in Sec.~\ref{sec:method}. This includes necessary details for the calculation of top pair production and six-fermion decay when mediated by a $Z'$ with full tree-level SM interference and all intermediate particles allowed off-shell (Sec.~\ref{sec:cross_section}), as well as those for the calculation of asymmetry observables (Secs.~\ref{sec:charge_asymmetry}, \ref{sec:top_polarisation_asymmetry}). The method also outlines the procedure employed in event generation (Sec~\ref{sec:event_generation}), and top reconstruction for the lepton-plus-jets final state where some experimental conditions arising for this channel are emulated. These include kinematic requirements and top quark pair reconstruction in the presence of missing transverse energy and combinatorial ambiguity in jet-top assignment, while remaining limited to the parton level (Sec.~\ref{sec:top_reconstruction}). We aim to assess the prospect for an LHC analysis to profile a $Z'$ boson mediating $t\bar{t}$ production, using both a standard bump-hunt via the cross section, as well as the charge asymmetry of the top quark system and the single top polarisation, with results and conclusions in Sec.~\ref{sec:results} and~\ref{sec:conclusions}, respectively.

\section{Models}
\label{sec:models}

\begin{figure}
  \centering
  \includegraphics[width=\linewidth]{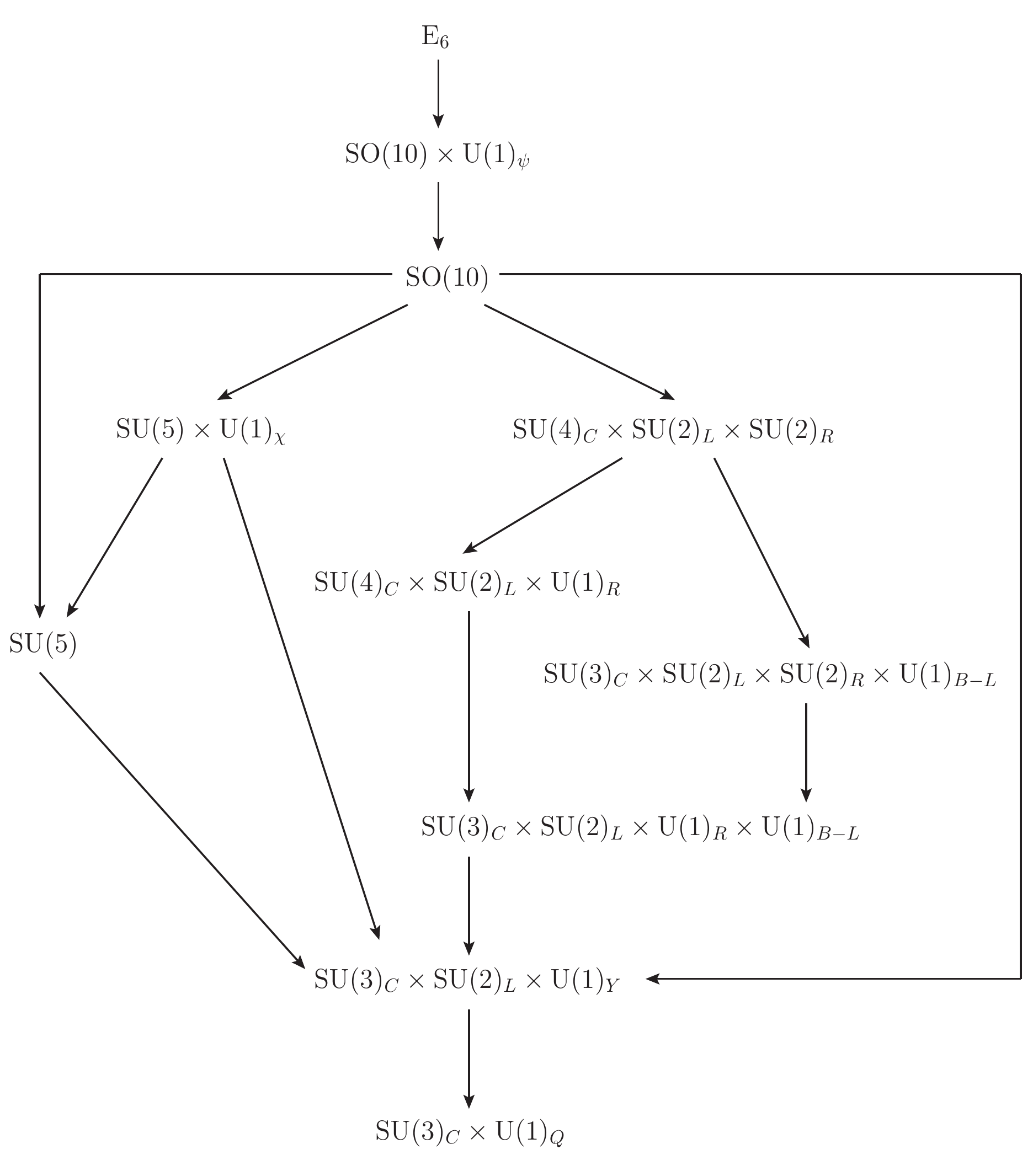}
  \caption{Chart of the possible spontaneous symmetry breaking chains featuring GUT candidates SU$(5)$, SO$(10)$ and E$_6$. The residual U$(1)$ symmetries can lead to the $Z'$ benchmark models explored in this study. Diagram based on a similar illustration in Ref.~\cite{diluzio2011}.}
  \label{fig:symmetry_breaking_chart}
\end{figure}

There are several candidates for a GUT, a hypothetical enlarged gauge symmetry, motivated by approximate gauge coupling unification at around the $10^{13}$~TeV energy scale~\cite{georgi1974,fritzsch1975,gursey1976,buras1978}. At lower energies these may proceed through sequential spontaneous symmetry breaking mechanisms, often leaving residual U$(1)$ gauge symmetries, until the familiar ${\rm SU}(3)\times {\rm SU}(2) \times {\rm U}(1)$ gauge structure of the SM is recovered. Examples of these cascade mechanisms are illustrated in Fig.~\ref{fig:symmetry_breaking_chart} featuring SO$(10)$ and E$_6$, which motivate a variety of models featuring a $Z'$.

In isolation there is no particular reason for these $Z'$ to have masses as low as the TeV scale; however, for supersymmetric implementations of these unified models, the extra U$(1)$ breaking scale is generally linked to the scale of soft supersymmetry breaking~\cite{cvetic1998}. Hence, the motivations for a TeV scale $Z'$ are the same as those for TeV scale supersymmetry, namely, a solution to the Hierarchy problem, and naturally occurring exact gauge coupling unification~\cite{feng2013}.

In each of the models, the residual U$(1)$ gauge symmetry is assumed to be broken around the TeV scale, resulting in a massive $Z'$ boson. This leads to an additional term in the low-energy Neutral-Current (NC) Lagrangian:
\begin{align}
  -\mathcal{L}'_{NC} 
                   & = e J^{\mu}_e A_{\mu} + g_Z J^\mu_Z Z_\mu + g'J'^\mu Z'_\mu,
\end{align}
for the electromagnetic coupling $e$ to the photon field $A^\mu$, $g_Z$ coupling to the $Z$ boson field $Z^\mu$, and $g'$ coupling to the $Z'$ boson field $Z'^\mu$. The later couples to a vector current:
\begin{equation}
  J'^{\mu} = \sum_f \bar{\psi}_f\gamma^\mu Q' \psi_f = \sum_f \bar{\psi}_f \gamma^\mu\left[g'^f_L P_L + g'^f_R P_R\right]\psi_f = \sum_f \bar{\psi}_f \gamma^\mu\frac{1}{2}\left[g'^f_V - g'^f_A\gamma_5\right]\psi_f,
  \label{eq:zprime_lagrangian}
\end{equation}
where $g'^f_{L(R)}$ are the chiral couplings to the left(right)-handed projections of a specific fermion field $P_L(P_R)\psi_f$, and $g'^f_{V,A}$ are the corresponding vector and axial couplings. These depend on the particular combination of generators in $Q'$, which along with the overall coupling $g'$, depend on the model. The charge generators of these couplings $Q'$ for each model class correspond to those defined in Eqs.~\ref{eq:QGSM}, \ref{eq:QGLR} and \ref{eq:QE6}. This term leads to a corresponding additional Feynman rule and diagram (Fig.~\ref{fig:lottdiagrams}c).

An assortment of benchmark examples of such models are studied here, particularly those found in Ref.~\cite{accomando2011}. These may be classified into three types: General Sequential Models (GSM), Generalised Left-Right symmetric (GLR) models and $E_6$ inspired models. Below follows a brief overview of each classification, followed by a summary of the assumptions and parameters used, as well as the present limits from the LHC.

\subsection{Generalised Sequential Models}
\label{sec:generalised_sequential_models}

Above the scale of EW symmetry breaking, the SM NC Lagrangian may be written as
\begin{equation}
  -\mathcal{L}^{SM}_{NC} = g_L J^{3\mu}_{L}W^3_{L\mu} + g_Y J_Y^\mu B_{Y\mu},
\end{equation}
where $g_L$ and $g_Y$ are the SU$(2)_L$ and U$(1)_Y$ gauge couplings, respectively. In the Sequential Standard Model (SSM) an additional neutral gauge boson is introduced with fermionic couplings generated identically to those of the SM $Z$ boson:
\begin{equation}
  g' Q'_{SSM} = g_Z Q_Z = \frac{g_L}{\cos\theta_W}\left(T^{3}_{L} - \sin^2\theta_W Q_e\right),
\end{equation}
where the weak mixing angle $\theta_W$ is defined by $\tan\theta_W = g_Y / g_L$.\footnotemark{} The fermionic couplings of this $Z'_{SSM}$ are uniquely determined by well defined eigenvalues of the $T^3_L$ and $Q_e$ generators, the third component of weak isospin and the Electro-Magnetic (EM) charge, respectively. This $Z'$ differs only in being generically heavier than the $Z$ boson and, consequently, by having a larger decay width (Eq.~\ref{eq:zprime_decay_width}).

\footnotetext{$\cos\theta_W = m_{W} / m_{Z}$ defines the weak mixing angle experimentally.}

The GSMs allow for a general linear combination of these generators:
\begin{equation}
  Q'_{\rm GSM} = \cos\theta' T^3_L + \sin\theta' Q_e.
  \label{eq:QGSM}
\end{equation}
The $Z'$ coupling is fixed by the condition that $g_Z Q_Z \equiv g'Q'_{\rm GSM}$ for a particular value of $\theta'$, implying that the SSM case is recovered when
\begin{equation}
 g' = \frac{g_L}{\cos\theta_W}\sqrt{1 + \sin^4\theta_W} \approx 0.76 \qquad {\rm and} \qquad \theta' = -0.072\pi.
\end{equation}
New models are generated keeping $g'$ fixed and varying $\theta'$ over its range $- \frac{\pi}{2} \le \theta' \leq \frac{\pi}{2}$. Clearly, at its extremes pure $T^3_L$ (left-chiral coupling) and $Q_e$ (heavy photon) models are recovered. The resultant couplings to fermions are given in Tab.~\ref{tab:zprime_parameters}.

\subsection{Generalised Left-Right symmetric models}

The LR symmetric model introduces a new isospin group SU$(2)_R$ perfectly analogous to the SU$(2)_L$ weak isospin group of the SM, but which acts on right-handed fields. The SU$(2)_R$ symmetry may then be broken to U$(1)_{R}$, linked to the generator $T^{3}_{R}$ associated to the third (diagonal) component of right isospin with eigenvalues $+\frac{1}{2}$ for $(u_R, \nu_R)$ and $-\frac{1}{2}$ for $(d_R, \nu_R)$:
\begin{equation}
   {\rm SU}(2)_L \times {\rm SU}(2)_R \times {\rm U}(1)_{B-L} \rightarrow
   {\rm SU}(2)_L \times {\rm U}(1)_R \times {\rm U}(1)_{B-L}.
\end{equation}
Here U$(1)_{B-L}$ relates to the generator $T_{BL} = \frac{1}{2}(B - L)$ and conserves the difference between the Baryon (B) and Lepton (L) numbers.

Furthermore, the gauge symmetry
\begin{equation}
   {\rm SU}(3)_C \times {\rm SU}(2)_L \times {\rm U}(1)_R \times {\rm U}(1)_{B-L}
\end{equation}
arises naturally in the course of breaking an SO$(10)$ GUT, either via SU$(2)_{R}$ or directly from a larger intermediate gauge symmetry (Fig.~\ref{fig:symmetry_breaking_chart}), leading to a NC Lagrangian
\begin{equation}
  -\mathcal{L}^{LR}_{NC} = g_L J^{3\mu}_{L}W^3_{L\mu} + g_R J^{3\mu}_{R}W^3_{R\mu} + g_{BL} J_{BL}^\mu B_{BL\mu}.
\end{equation}
All of these versions are anomaly-free after including the three right-handed neutrinos $\nu_R$~\cite{langacker2009}.\footnotemark{}

To reproduce the standard model, the hypercharge preserving symmetry must be recovered:
\begin{equation}
  {\rm U}(1)_R \times {\rm U}(1)_{B-L}\rightarrow {\rm U}(1)_Y,
\end{equation}
at a scale $m_{Z'} >> m_Z$, making the hypercharge generator $T_Y = T^3_{R} + T_{BL}$, with the mass eigenstates of the fields described by orthogonal relations
\begin{align}
  B^\mu_Y  & = \cos\phi W^{3\mu}_R + \sin\phi B_{BL}^\mu, \\
  Z'^\mu_{LR} & = - \sin\phi W^{3\mu}_R + \cos\phi B_{BL}^\mu.
\end{align}
Here $\tan\phi = g_R/g_{BL}$ and $B^\mu_Y$ is the massless boson of the U$(1)_Y$ symmetry, such that its gauge coupling constant is defined through the relation $1/g_Y^2 = 1/g_R^2 + 1/g_{BL}^2$.

Therefore, equating the coupling $g'_{LR}$ with the GUT normalised $g_Y$ of the SM. $Z'^\mu_{LR}$ couples to a current associated with the charge:
\begin{align}
  Q_{\rm LR}' = & \sqrt{\frac{5}{3}} \left(\tan\phi T^3_R - \cot\phi T_{BL} \right),\\
  \tan\phi = & \sqrt{\frac{g_R^2}{g_L^2}\cot^2\theta_W - 1}, \\
  g_{Y} = & \sqrt{\frac{3}{5}}g_L\tan\theta_W \approx 0.46.
\end{align}

Furthermore, Generalised LR (GLR) symmetric models may be considered, in which the $Z'$ corresponds to a general linear combination of U$(1)_R$ and U$(1)_{B-L}$,
\begin{equation}
  Q'_{\rm GLR} = \cos\theta' T^3_R + \sin\theta' T_{BL},
  \label{eq:QGLR}
\end{equation}
The gauge coupling $g'$ is fixed by $g'_{Y} Q'_{\rm LR}\equiv g'Q'_{\rm GLR}$ for a specific $\theta'$ so that the original LR model may be recovered by setting $g' = 0.595$ and $\theta' = -0.128\pi$. Three further special cases may be identified. Obviously, taking $\theta' = 0, \pi/2$ produces purely $T^3_R$, $T_{BL}$ generated resonances, respectively, while $\theta' = \pi/4$ leads to a $Z'$ which couples directly to hypercharge (Tab.~\ref{tab:zprime_parameters}).

\subsection{E\texorpdfstring{$_6$}{6} inspired models}

One may propose that the gauge symmetry group at the GUT scale is E$_6$. As illustrated in Fig.~\ref{fig:symmetry_breaking_chart}, this proceeds via a series of symmetry breaking mechanisms until the SM gauge symmetry is recovered. For this study, the important mechanisms are those that lead to an extra residual U$(1)$, namely,
\begin{align}
  {\rm E}_6 &\rightarrow {\rm SO}(10)\times {\rm U}(1)_\psi,\\
  {\rm SO}(10) &\rightarrow {\rm SU}(5)\times{\rm U}(1)_\chi, \\
  {\rm SU}(5) &\rightarrow {\rm SU}(3)_C \times {\rm SU}(2)_L \times {\rm U}(1)_Y.
\end{align}

All of these may occur around the GUT scale, yet it is possible for the U$(1)_\psi$ and U$(1)_\chi$ to survive down to the TeV scale. Therefore, a general linear combination of these generators may be taken,
\begin{equation}
  Q_{E_6}=\cos\theta' T_\chi + \sin \theta' T_\psi.
  \label{eq:QE6}
\end{equation}

Six special cases of E$_6$ inspired $Z'$ models are considered, with couplings to quarks $g'Q_{E_6}$. The $g'$ value is again equated to the GUT normalised hypercharge coupling of the SM. Further considerations of the GUT breaking pattern and normalisation conditions lead to the quark couplings for each case and are shown in Tab.~\ref{tab:zprime_parameters}. The $\eta$ model occurs in the Calabi-Yau compactifications of the heterotic string where E$_6$ breaks directly to a rank 5 group~\cite{witten1985}, while the inert ($I$) model has an orthogonal charge to that in the $\eta$ model via an alternative $E_6$ breaking pattern~\cite{robinett1982}. The $N$ model formulates a U$(1)$ symmetry in which $\nu_R$ have zero charge, allowing very heavy Majorana masses suitable to take part in a seesaw mechanism to yield small $\nu_L$ masses~\cite{king2006}. In the secluded sector model ($S$) the U$(1)$ is broken in a sector coupled minimally to ordinary fields~\cite{erler2002}.

\subsection{Summary and present limits}
\label{sec:summary_and_present_limits}

For each model group, $g'$ is fixed and the angular parameter dictating the relative strengths of the component generators varied until interesting limits for each class, as outlined above, are recovered. For the generalised models, in addition to the central values, the interesting limits simply maximise the contribution from either generator. The resulting parameters are summarised in Tab.~\ref{tab:zprime_parameters}.

\begin{table}
  \centering
  \begin{tabularx}{0.9\textwidth}{l c l c c c c c c}
    \toprule
    Class & $g'$ & Model & $\theta$ & $g'^u_V$  & $g'^u_A$   & $g'^d_{V}$  & $g'^d_{A}$ & $\Gamma_{Z'} / m_{Z'}$ (\%)  \\
    \midrule
    \multirow{3}{*}{GSM}   & \multirow{3}{*}{$0.760$} & $L$     & $0        $ & $0.5    $ & $0.5    $ & $-0.5  $ & $-0.5 $ & $4.7$ \\
                           &                          & $Q$     & $0.5\pi   $ & $1.333  $ & $0      $ & $-0.666$ & $0    $ & $12.5$\\
                           &                          & $SSM$   & $-0.072\pi$ & $0.193  $ & $0.5    $ & $-0.347$ & $-0.5 $ & $3.2$ \\
    \midrule
    \multirow{4}{*}{GLR}   & \multirow{4}{*}{$0.595$} & $R$     & $0        $ & $0.5    $ & $-0.5   $ & $-0.5  $ & $0.5  $ & $2.5$ \\
                           &                          & $BL$    & $0.5\pi   $ & $0.333  $ & $0      $ & $-0.333$ & $0    $ & $1.5$ \\
                           &                          & $LR$    & $-0.128\pi$ & $0.329  $ & $-0.46  $ & $-0.591$ & $0.46 $ & $2.1$ \\
                           &                          & $Y$     & $0.25\pi  $ & $0.589  $ & $-0.353 $ & $-0.118$ & $0.354$ & $2.4$ \\
    \midrule
    \multirow{6}{*}{E$_6$} & \multirow{6}{*}{$0.462$} & $\chi$  & $0       $  & $0$       & $-0.316 $ & $-0.632$ & $0.316$ & $1.2$ \\
                           &                          & $\psi$  & $0.5\pi  $  & $0$       & $ 0.408 $ & $0     $ & $0.408$ & $0.5$ \\
                           &                          & $\eta$  & $-0.29\pi$  & $0$       & $-0.516 $ & $-0.387$ & $0.129$ & $0.6$ \\
                           &                          & $S$     & $0.129\pi$  & $0$       & $-0.129 $ & $-0.581$ & $0.452$ & $1.2$ \\
                           &                          & $I$     & $0.21\pi $  & $0$       & $ 0     $ & $ 0.5  $ & $-0.5 $ & $1.1$ \\
                           &                          & $N$     & $0.42\pi $  & $0$       & $-0.316 $ & $-0.158$ & $0.474$ & $0.6$ \\
    \bottomrule
  \end{tabularx}
  \caption{Benchmark model $Z'$ parameters and the resultant axial-vector couplings to quarks, in addition to the decay-width to resonant-mass ratio as a percentage~\cite{accomando2011}.}
  \label{tab:zprime_parameters}
\end{table}

These models are all universal, with the same coupling strength to each generation of fermion. Therefore, as with an SSM $Z'$, the strongest experimental limits come from the DY channel. Analysing $20.3$~fb$^{-1}$ of proton-proton data at $\sqrt{s}=8$~TeV from the LHC, both the ATLAS experiment and CMS experiment exclude a Sequential SM (SSM) $Z'$, at the $95\%$~confidence level, for masses lower than $2.90$~TeV, in the combined electron and muon channels~\cite{theatlascollaboration2014c,thecmscollaboration2015}. Based on these results, the limits for the GUT inspired models have been extracted in Ref.~\cite{accomando2016} (Tab.~\ref{tab:zprime_limits}). Using $3.2$~fb$^{-1}$ of $\sqrt{s} = 13$~TeV data, the ATLAS collaboration has published combined results in DY that place a lower limit on the mass of an SSM $Z'$ at $3.36$~TeV~\cite{theatlascollaboration2016b}, while presently unpublished results, from both collaborations, with approximately $13$~fb$^{-1}$ push this limit to $4$~TeV~\cite{theatlascollaboration2016a,thecmscollaboration2016}. As indicated in Tab~\ref{tab:zprime_limits}, mass limits for this assortment of models are generally close to those of an SSM $Z'$; $4$~TeV is selected as the benchmark mass for the new resonance.

The present limits coming from $t\bar{t}$ events are typically lower, with the most stringent limits on a narrow leptophobic $Z'$ excluding masses only less than $2.4$~TeV~\cite{atlascollaboration2015, khachatryan2016}. Therefore, when studying non-universally coupling models (with enhanced third generation couplings), those which embed one or more $Z'$ with lower masses than generally prohibited by DY results may be explored.

\vskip8mm

\begin{table}
  \setlength{\tabcolsep}{5pt}
  \centering
  \begin{tabularx}{\textwidth}{l|ccccccccccccc}
    Model          & $\chi$ & $\psi$ & $\eta$ & $S$    & $I$    & $N$    & $R$    & ${BL}$ & ${LR}$ & $Y$    & ${SSM}$ & ${T^3_L}$ & ${Q}$ \\
    \hline
    $m_{Z'}$ [TeV] & $2.70$ & $2.56$ & $2.62$ & $2.64$ & $2.60$ & $2.57$ & $3.04$ & $2.95$ & $2.77$ & $3.26$ & $2.90$  & $3.14$    & $3.72$ \\
  \end{tabularx}
  \caption{Lower bounds on the mass of a new resonance ($m_{Z'}$) embedded by benchmark GUT models, based on $\sqrt{s} = 8$~TeV CMS results with an integrated luminosity of $L = 20.3$~fb$^{-1}$~\cite{thecmscollaboration2015, accomando2016}.}
  \label{tab:zprime_limits}
\end{table}

\section{Method}
\label{sec:method}

\subsection{Cross section}
\label{sec:cross_section}

Due to the differing initial state, the QCD diagrams in Fig.~\ref{fig:lottdiagrams}a and \ref{fig:lottdiagrams}b do not interfere with those of \ref{fig:lottdiagrams}c and the square matrix element may be linearly separated as
\begin{equation}
  |\mathcal{M}(pp\rightarrow t\bar{t})|^2 = |\mathcal{M}({\rm QCD})|^2 + |\mathcal{M}({\rm NC}\{\gamma, Z, Z'\})|^2.
\end{equation}
Hence, the $Z'$ interferes only with the neutral EW bosons. Using stable top quarks for illustrative simplicity, the total square matrix element may be written as
\begin{equation}
    |\mathcal{M}(\gamma, Z, Z')|^2
    = \frac{\hat{s}^2}{6}\frac{D^{ij}}{1+\delta^{ij}}\left\{C_q^{ij}\left[C_t^{ij}(1+\beta^2\cos^2\theta) +
    B_t^{ij}(1-\beta^2)\right] + 2A_q^{ij}A_t^{ij}\beta \cos \theta\right\}.
    \label{eq:Mqqtt2}
\end{equation}
Here, $q$ denotes the initial quark species, coming from the proton, while $t$ denotes the final top quark. The labels $i$ and $j$ denote the vector bosons $\{\gamma, Z, Z'\}$ and repeated indices are summed over. The angle $\theta$ is between the top quark momentum and the incoming quark direction, in the $t\bar{t}$ centre of mass frame. For the chiral couplings $g_{L,R}$ of a particular fermion $f$ define:
\begin{align*}
  A_{f}^{ij} & \equiv g^{if}_{L} g^{jf}_{L} - g^{jf}_{R} g^{if}_{R}, \\
  B_{f}^{ij} & \equiv g^{if}_{L} g^{jf}_{R} + g^{jf}_{R} g^{if}_{L}, \\
  C_{f}^{ij} & \equiv g^{if}_{L} g^{jf}_{L} + g^{jf}_{R} g^{if}_{R}, \\
  D_{ij} & \equiv \frac{(\hat{s}-m_i^2)(\hat{s}-m_j^2) + m_im_j\Gamma_i\Gamma_j}{\left((\hat{s}-m_i^2)^2 +m_i^2\Gamma^2_i\right)\left((\hat{s}-m_j^2)^2 + m_j^2 \Gamma_j^2 \right)}.
\end{align*}

Eq.~\ref{eq:Mqqtt2} illustrates the potential for interference between the SM vector bosons and new resonances.
Isolating the $Z'$ only element and integrating over $\cos\theta$, the coupling structure for the cross section, in vector and axial-vector notation is:
\begin{equation}
  \hat{\sigma} \propto \left({g'^q_V}^2 + {g'^q_A}^2\right)\left((4 - \beta^2){g'^t_V}^2 + {g'^t_A}^2\right).
  \label{eq:sigma_coupling}
\end{equation}

The $Z'$ is assumed to decay only to SM fermions.\footnotemark{} Thus, the decay width of the $Z'$ is
\begin{equation}
	\Gamma(Z' \rightarrow f\bar{f}) = \sum_f \frac{g^2_{Z'}m_{Z'}}{48\pi}\beta\left[\frac{3-\beta^2}{2}{g'^f_V}^2 + \beta^2 {g'^f_A}^2\right],
  \label{eq:zprime_decay_width}
\end{equation}

where $\beta \equiv \sqrt{1 - 4\frac{m_f^2}{m_{Z'}}}$.

\footnotetext{Though here unconsidered, if the $Z'$ is allowed to mix with the SM $Z$ it can lead to two additional decay modes: $Z'->W^{+}W^{-}$ and $Z'\rightarrow hZ$.}

\subsection{Charge asymmetry}
\label{sec:charge_asymmetry}

Charge asymmetries refer to the symmetry of a process under charge conjugation. For a charge-parity preserving process, this equates to an asymmetry in the spatial or angular dependence of the two body decay products. The simplest example of a spatial asymmetry is the forward-backward asymmetry ($A_{FB}$). The detector region is split into two hemispheres and the number of top quarks, or antitop quarks, detected on either side are compared:
\begin{equation}
	A^t_{FB} = \frac{N_{t}(\cos\theta > 0) - N_t(\cos\theta < 0)}{N_t(\cos\theta > 0) + N_t(\cos\theta < 0)},
\end{equation}
where the angle $\theta$ is between the top quark momentum and the incoming quark direction, in the centre of mass frame of the incoming pair. This asymmetry separates the cross section by integrating over opposite halves of the angular phase space:
\begin{equation}
	\hat{\sigma}_F = \int_0^1 \frac{d\hat{\sigma}}{d\cos\theta}d\cos\theta, \qquad	\hat{\sigma}_B = \int_{-1}^0 \frac{d\hat{\sigma}}{d\cos\theta}d\cos\theta.
\end{equation}
Retaining the angular dependence of the matrix element with full interference between $\gamma$, $Z$ and $Z'$ (see Eq.~\ref{eq:Mqqtt2}), only the last term survives the subtraction in the $A_{FB}$ numerator. Hence,
\begin{equation}
	A^t_{FB} \propto A_g^{ij}A_t^{ij}.
\end{equation}
Examining the $Z'$ in isolation gives, in fact:
\begin{equation}
	A^t_{FB} \propto g'^q_V g'^q_A g'^t_V g'^t_A,
  \label{eq:afb_coupling}
\end{equation}
which is uniquely different from the coupling structure of $\hat{\sigma}$ (Eq.~\ref{eq:sigma_coupling}).

With hadrons in the initial state though, the quark direction is indeterminate. However, $A_{FB}$ has been very useful when using $p\bar{p}$ colliders, such as the Tevatron, where (recalling that the $Z'$ process is induced by $q\bar q$ initial states) the $z$ direction is highly correlated with the incoming $q$ direction, as the contribution due to an interaction between a sea $q$ from the $\bar{p}$ and a sea $\bar{q}$ from the $p$ is very small.

At the LHC, however, the protons have identical parton density functions. Hence, the forward and backward direction do not correlate with the most probable quark direction. However,  the $q$ is likely to have a larger partonic momentum fraction $x$ than the $\bar{q}$ in $\bar{x}$. Therefore, choose the $z^*$ axis to lie along the boost direction. Thus the cosine of the angle measured with respect to the collider $z$ axis is multiplied by the sign of the rapidity of the top quark pair in the collider frame $y_{tt}$:
\begin{equation}
	\cos\theta^* = \frac{y_{tt}}{|y_{tt}|}\cos\theta.
\end{equation}
Thus a new variable may be defined, known as the reconstructed forward-backward asymmetry ($A^{t}_{FB^{*}}$):
\begin{equation}
	A^{t}_{FB^{*}} = \frac{N_{t}(\cos\theta^* > 0) - N_t(\cos\theta^* < 0)}{N_t(\cos\theta^* > 0) + N_t(\cos\theta^* < 0)}.
\end{equation}
It is not immediately apparent, but $A^{t}_{FB^{*}}$ is effectively equivalent to another common definition of charge asymmetry, $A_C$:
\begin{equation}
	A_{C} = \frac{N_{t}(\Delta|y| > 0) - N_t(\Delta|y| < 0)}{N_t(\Delta|y| > 0) + N_t(\Delta|y| < 0)}.
\end{equation}
This definition is used to measure the charge asymmetry by the LHC collaborations. In some literature it is called $A_{RFB}$, for rapidity-dependent forward-backward asymmetry~\cite{basso2012a}.

\subsection{Top polarisation asymmetry}
\label{sec:top_polarisation_asymmetry}

The top polarisation asymmetry, or single spin asymmetry ($A_{L}$), measures the net polarisation of the (anti)top quark by subtracting events with positive and negative helicities. For the top quark, it is defined
\begin{equation}
  A_{L} = \frac{N(+,+) + N(+,-) - N(-,-) - N(-,+)}{N(+,+) + N(+,-) + N(-,-) + N(-,+)},
\end{equation}
where $N(\lambda_{t},\lambda_{\bar{t}})$ denotes the number of events observed with eigenvalues of the helicity operator $\lambda_t=\pm$ and $\lambda_{\bar{t}}=\pm$ for the top quark and antitop quark, respectively.

In order to determine the coupling structure for the numerator of $A_L$, it is necessary to first calculate the polarised matrix elements of $Z'$ production and decay, which has been done and verified against Ref.~\cite{arai2009}. This is most straightforward using the left-right couplings. With these, and suppressing the propagator factor, the matrix element terms that survive in the numerator can be shown to have the form
\begin{align}
	\mathcal{M}_{A_{L}} = \frac{\hat{s}^2}{12}\left\{(q_L^2 + q_R^2)(t_L^2 - t^2_R)(1 + \cos^2\theta) + 2(q_L^2 - q^2_R)(t^2_L + t^2_R)\cos\theta\right\}.
\end{align}
This may be generalised to include multiple neutral bosons, providing the coupling structure
\begin{equation}
	A_L \propto \beta C_q^{ij} A_t^{ij}.
\end{equation}
Once again, examining the $Z'$ in isolation:
\begin{equation}
  A_{L} \propto \left({g^q_V}^2 + {q^t_A}^2\right)q^t_V g^t_A \beta.
  \label{eq:al_coupling}
\end{equation}
Interestingly, the incoming quark dependence is proportional to the square of the couplings, while the dependence on the top couplings is linear. This observable is, therefore, directly sensitive to the chirality of the $Z'$ coupling to the top~\cite{bernreuther2006,moretti2006}.

Spin asymmetries are not directly observable at the LHC.\footnotemark{} However, given to the short life time of top quarks they decay prior to hadronisation. Information on the spin is, therefore, retained by the decay products. Hence, the angular distributions of these products are sensitive to the spin configuration.

\footnotetext{There are two principal spin asymmetries for top pair production; the spin correlation asymmetry
\begin{equation}
  A_{LL} = \frac{N(+,+) + N(-,-) - N(+,-) - N(-,+)}{N(+,+) + N(-,-) + N(+,-) + N(-,+)},
\end{equation}
is best measured in the dilepton final state. Furthermore, its coupling dependence
\begin{equation}
  A_{LL} \propto \left({g'^q_V}^2 + {g'^q_A}^2\right)\left((2 + \beta^2){g'^t_V}^2 + 3{g'^t_A}^2\right)
  \label{eq:all_coupling}
\end{equation}
is very similar to the cross section itself, therefore, it does not offer an additional handle for profiling of $Z'$. Accordingly, its potential is not explored in this study.}

Treating the top quark in the narrow width approximation, the total partonic matrix element squared for top pair production, including decay channels, is
\begin{equation}
  |\mathcal{M}|^2 \propto {\rm Tr}[\bar{\rho} R \rho] = \bar{\rho}_{\lambda_{\bar{t}}'\lambda_{\bar{t}}} R_{\lambda_{\bar{t}}\lambda_{\bar{t}}',\lambda_t\lambda_t'} \rho_{\lambda_t\lambda_t'},
  \label{one}
\end{equation}
where $\lambda_{(\bar{t})t}$ denotes the spin of the (anti)top and $R$ represents the density matrix for on-shell top pair production from initial partons, averaged over initial spins. Here, ($\bar{\rho}$)$\rho$ is the density matrix corresponding to the decay of the polarised (anti)top:
\begin{equation}
  \rho_{\lambda_t \lambda_t'} = \frac{\Gamma_f}{2}\left(1 + \kappa_f\boldsymbol{\sigma} \cdot \textbf{q}_f\right)_{\lambda_t \lambda_t'},
  \label{eq:top_polarisation_1}
\end{equation}
where $\boldsymbol{\sigma}$ are the Pauli matrices, and $\kappa_f$ is the `spin analysing power' of the particular decay product $f$ with decay width $\Gamma_f$ and 3-momentum $\textbf{q}_f$. Information about the top spin is preserved in the distribution of $\cos\theta_f$, the angle between the top momentum in the $t\bar{t}$ rest frame and the momentum of the decay fermion $f$ in the rest frame of the parent top. In the rest frame of the top, this distribution has the form:
\begin{equation}
  \frac{1}{\Gamma_f}\frac{d\Gamma_f}{dcos\theta_f} = \frac{1}{2}(1 + \kappa_f P_t \cos\theta_f),
	\label{eq:top_polarisation_2}
\end{equation}
where $P_t$ is the polarisation of the ensemble. The product $\kappa_f P_t$ is equivalent to the $A_L$ observable defined for stable polarised top quarks. The spin analysing powers at tree level for different decay product species are shown in Tab.~\ref{tab:spin_analysing_power}.

\begin{table}
	\centering
	\begin{tabularx}{0.4\textwidth}{c|cccccc}
	  $f$ & $l$ & $d$ & $\nu$ & $u$ & $W$ & $b$ \\
	  \hline
	  $\kappa_f$ & 1 & 1 & 0.3 & 0.3 & 0.39 & -0.39 \\
	\end{tabularx}
	\caption{Spin analysing powers of $t\bar{t}$ decay products.}
  \label{tab:spin_analysing_power}
\end{table}

The relationship outlined above enables $A_L$ to be probed using the angular distribution of the decay lepton. To compute $A_{L}$ binned in the invariant mass of the $t\bar{t}$ pair ($m_{tt}$); two dimensional histograms in $m_{tt}$ and $(\cos\theta_{\ell})$ are constructed. As described in Sec.~\ref{sec:top_reconstruction}, events are generated in which either the top quark or antitop quark decays leptonically. As $A_L$ is best measured using the decay lepton, when the (anti)top decays, i.e. an $\ell^-$($\ell^+$) is detected in an event, bins are made in $\cos\theta_{\ell^-(\ell^+)}$, for a combined histogram, binned in $m_{tt}$ and $\cos\theta_{\ell}$. Each mass slice is normalised by the integral of that slice, multiplied by two (due to the factor in Eq.~\ref{eq:top_polarisation_2}) and divided by the $\cos\theta_{\ell}$ bin width. For each mass slice a straight line is fitted to the $\cos\theta_{\ell}$ distribution. $A_L$ is extracted from the gradient of the fit to $\cos\theta_{\ell}$.

\subsection{Event generation}
\label{sec:event_generation}

The generation tool employed for our study is a custom Monte Carlo (MC) simulation program. The matrix element calculations are made to leading order (LO) based on helicity amplitudes using HELAS subroutines, with default Standard Model square matrix elements built up using MadGraph~\cite{hagiwara1992,stelzer1994}. Beyond the Standard Model amplitudes are then constructed by duplicating and modifying the matrix element code, as required. The VEGAS Amplified (VAMP) package is used for the multi-dimensional numerical phase-space integration~\cite{lepage1978,ohl1999}. The output is written directly to partonic ROOT $n$-tuples containing only the final state 4-momenta, particle species ID and event weight, using RootTuple~\cite{hall2012}.\footnotemark{}

\footnotetext{LHEF output is also available, in preparation for the extension to include the full parton-shower, hadronisation and detector simulation~\cite{alwall2007}.}

The CT14LL (LO) table is selected for these simulations, with a factorisation/renormalisation scale of $Q = \mu = 2 m_t$~\cite{dulat2016}.\footnotemark{} The $b$ and $t$ quarks are assigned masses of $4.18$ GeV and $172.5$ GeV, respectively, while the lighter quarks are treated in the massless limit.

\footnotetext{A number of different PDF sets are available for use with tool. The most recent of these are the CT14 leading order tables; however, CTEQ6 and MRS99 sets are provided, particularly for verification with previous results~\cite{pumplin2002,lai2010,gao2014}.}

The results of this tool have been validated against Refs.~\cite{basso2012a} and~\cite{accomando2016} using the same parameter choices and, upon resolving a bug found in the former paper resulting in a factor of four increase in the calculation of the $Z'$ decay width, results agree perfectly.

In this paper, all events are generated at LO without higher-order corrections. The Next-to-Leading Order (NLO) QCD corrections to the SM $t\bar{t}$ cross-section have been developed extensively, proceeding from stable, on-shell top quarks~\cite{nason1989,beenakker1991,mangano1992,frixione1995} to those that account for their decay using a spin-correlated narrow-width approximation~\cite{bernreuther2004a,melnikov2009,bernreuther2010} to fully off-shell calculations~\cite{denner2011,denner2012,denner2012a,bevilacqua2011}. The NLO EW corrections have also been studied comprehensively for stable top quarks~\cite{beenakker1994,moretti2006,bernreuther2006,kuhn2007,bernreuther2008a}. More recently NNLO QCD corrections have been fully calculated~\cite{barnreuther2012,czakon2015,czakon2016}, as well Next-to-Next-to-Leading Log (NNLL) threshold resummation of soft radiation~\cite{czakon2009,czakon2009a,beneke2010}. The NLO QCD corrections to a new neutral heavy resonance have also been calculated in Refs.~\cite{gao2010,bonciani2016}, where it is found that the total cross section for $Z'\rightarrow t\bar{t}$ can be enhanced by a $K$-factor of $1.2 - 1.4$, depending on the mass of the resonance; however, the shape of the NLO distributions are not significantly different from those at LO, with a negligible effect on the spin correlations of the top quark pair.\footnote{A Randall-Sundrum KK Graviton can have a total NLO $K$-factor as high as $2.0$, with distributions significantly different from those at LO.}

The focus of this paper is on the asymmetry observables $A^t_{FB}$ and $A_L$, and their use in distinguishing model classes, and supplementing the cross section as a complementary discovery observable. At LO $A^t_{FB}$ and $A_{L}$ are zero and relatively flat as a function of $m_{tt}$, respectively.\footnotemark{} Both experiment and NNLO predictions seem to show a linearly increasing $A_{FB}$ with $m_{tt}$ between $350$~GeV and $750$~GeV~\cite{d0collaboration2014,czakon2015}; however this trend may not be extrapolated to higher energies. Indeed LHC measurements with $\sqrt{s} = 8$~TeV data show a negligible charge asymmetry over the full energy range~\cite{cmscollaboration2016,atlascollaboration2015a,atlascollaboration2016a,atlascollaboration2016b}. Likewise NLO predictions of top polarisation indicate negligible top polarisation~\cite{moretti2006}, which persists experimentally~\cite{atlascollaboration2013c}. Therefore, the assumption may be made that NLO and NNLO contributions cancel in the asymmetries, and higher-loop effects may be safely disregarded for the purposes of this study.

\footnotetext{In the SM there is a non-zero LO contribution to $A_{L}$ from the $Z$ boson.}

\subsection{Top reconstruction}
\label{sec:top_reconstruction}

\begin{figure}
  \centering
  \includegraphics[width=0.6\textwidth]{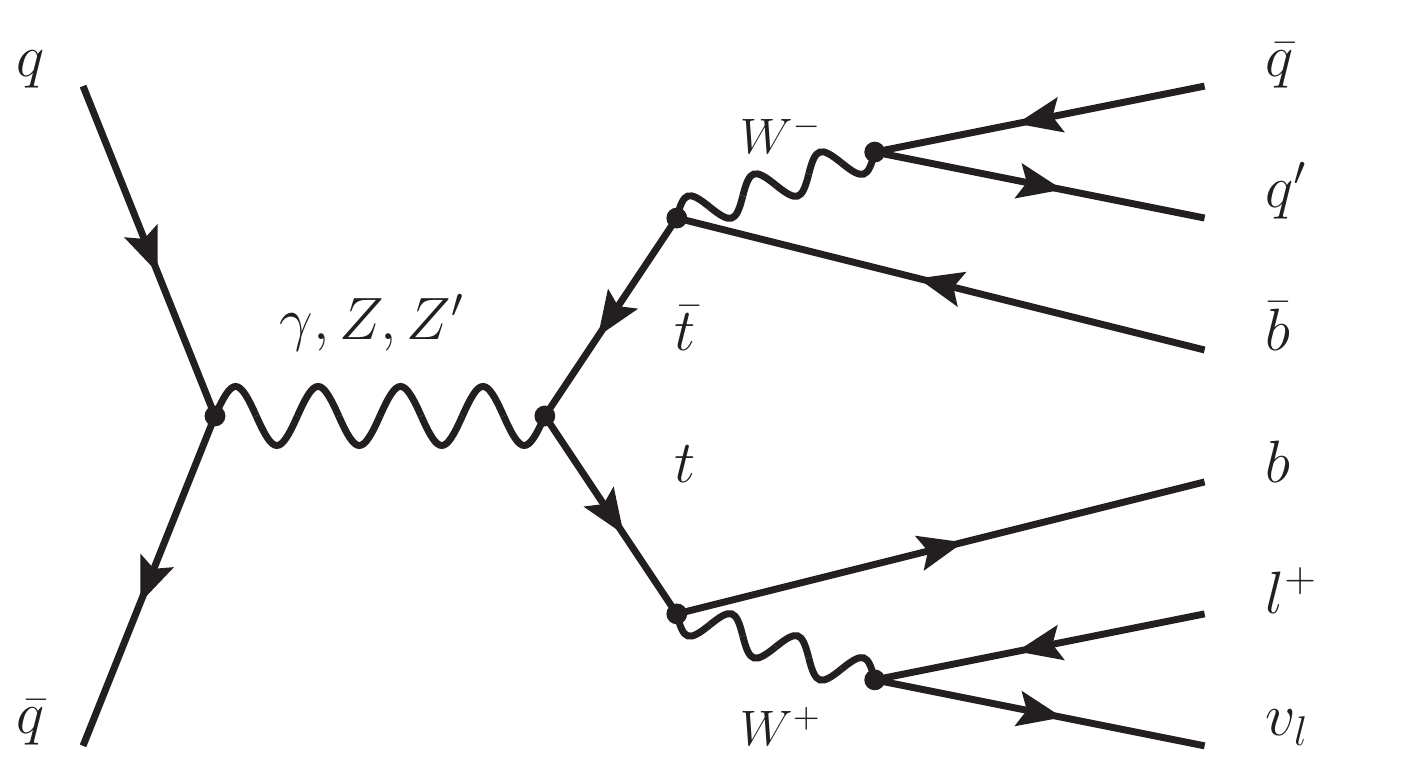}
  \caption{Feynman diagram for the signal process of LO lepton-plus-jets top pair production mediated by an interfering photon, $Z$ boson and $Z'$ boson.}
  \label{fig:qqblvqq}
\end{figure}

As $|V_{tb}|\approx1$, the top decays almost exclusively via $t\rightarrow bW^+$, such that the final state objects are entirely determined by the decay of the $W^\pm$ boson.\footnotemark The $W^+(W^-)$ boson may decay either via $W^+(W^-)\rightarrow \ell^+\nu(\ell^-\bar{\nu})$ or $W^{\pm}\rightarrow q\bar{q}'$. This leads to three classifications of a $t\bar{t}$ event: if both $t$ and $\bar{t}$ decay via the latter process, this is known as the fully hadronic channel; if both decay via the former process, the channel is dileptonic; else, it is semileptonic, or, based on its collider signature, lepton-plus-jets. The lepton-plus-jets channel is the sole focus here, for the aforementioned reasons (Sec.~\ref{sec:introduction}). In the EW sector, the Feynman diagram for this process at tree-level is shown in Fig.~\ref{fig:qqblvqq}.

\footnotetext{Where $|V_{tb}|$ corresponds to the magnitude of the element from the CKM quark mixing matrix relevant in determining the probability of top quark decays to bottom quarks.}

In order to minimise dataset file size, the generation code is written to produce $2\rightarrow6$ top events in a generalised way, which can then be assigned a particular decay channel (dileptonic, lepton-plus-jets and fully hadronic) at the analysis stage. This can be done because, at parton-level, assuming only the bottom quark has a mass, the particle species of the decay products affect the cross section only up to an overall constant. This constant can then be applied at any point during the analysis stage.

Our study is currently carried out at the parton level; However, multiple restraints encountered in a genuine analysis performed on reconstructed data are incorporated in the analysis. The following considerations are meant to mimic realistic experimental conditions, in order to assess, in a preliminary way, whether these observables and techniques may be of value in an experimental analysis. The collider signature for our process is a single lepton produced with at least four jets, in addition to missing transverse energy, $E^{\rm miss}_{T}$. The lepton may be either an electron or a muon, while taus are disregarded.\footnotemark{}

\footnotetext{Tau leptons will decay leptonically approximately one third of the time, resulting in the same semilepton final state, but with an additional source of missing transverse energy with less energetic leptons. Unfortunately, this would disturb our proposed reconstruction, and this will distort the asymmetry distributions. This effect will be addressed in future work.}

On the hadronically decaying leg of the decay, the number of successful $b$-tags must be selected: zero, one or two. Experimentally, it is not possible to determine the charge of $b$-tagged jets. Therefore, there is ambiguity on which $b$-jet is associated with the parent top or anti-top, even with two $b$-tags. With two $b$-tags, there are $^2P_2 \times ^2C_1 = 2$ possible arrangements, with one there are $^2P_1 \times ^3P_1 \times ^2C_2 = 6$ while with zero one has $^4P_2 \times ^2C_2 = 12$.\footnotemark{} Hence, as $b$-tags are lost, the combinatorics increases rapidly. Of course, this analysis is further simplified as background processes or initial/final state radiation, which contribute additional jets, are unaccounted for. Jet charge could be used to reduce multiplicity in practice.

\footnotetext{The notation $^n P_r$ ($^n C_r$)denotes the permutation (combination) choosing $r$ solutions from $n$ possibilities.}

On the leptonically decaying leg of the decay (anti)top quark reconstruction must account for the invisible neutrino. As there is only a single source of $E_T^{\rm miss}$ in the final state, the transverse neutrino momentum may be identified with the $E_T^{\rm miss}$ of the final state in the collider frame, i.e., the negative sum of all $p_T$ for all the visible particles in final state ($p_i$):
\begin{equation}
  p^\nu_T=-\sum_{i} p_T^{i}.
\end{equation}
This then leaves a final unknown: the longitudinal component of the neutrino momentum $p^{\nu}_z$. Assuming an on-shell $W^\pm$ and both $e$ and $\mu$ to be massless,
\begin{equation}
  m_{W}^{2} = 2 p_{\nu} p_{e},
\end{equation}
where
\begin{align}
  p_\nu & = \left(\sqrt{{p^{\nu_T}}^2 + {p^{\nu_z}}^2},\textbf{p}^\nu_T, p_z^\nu\right), \\
  p_e & = \left(|p^e|, \textbf{p}^e_T, p_z^e\right).
\end{align}
Solving this equation leads to a quadratic one:
\begin{equation}
	{p_T^e}^2 {p_Z^\nu}^2 - 2kp_z^e p_z^\nu + {p_T^\nu}^2|p^e|^2 - k^2 = 0,
\end{equation}
where
\begin{equation}
  k = \frac{m_W^2}{2} + \textbf{p}_T^e \textbf{p}_T^\nu.
\end{equation}
Therefore, the approximate neutrino momentum may be calculated with a two-fold ambiguity. The solutions to the quadratic equation can be either wholly real or complex and there are a number of options for how to treat these. When the solutions are complex the event can be discarded, or the real part taken, which is identical for each solution. The approach taken in this study.

In order to reconstruct the event, bottom-top assignment and $p_z^\nu$ solution selection may be accounted for simultaneously, using a chi-square-like test, by minimising the variable $\chi^2$:
\begin{equation}
  \chi^2 = \left(\frac{m_{bl\nu} - m_{t}}{\Gamma_t}\right)^2 + \left(\frac{m_{bqq} - m_{t}}{\Gamma_t}\right)^2,
  \label{eq:chi2}
\end{equation}
where $m_{bl\nu}$ and $m_{bqq}$ are the invariant mass of the leptonic and hadronic (anti)top quark, respectively, while $m_t$ and $\Gamma_t$ are the top quark mass and decay width, respectively.

The analysis code is written within the ROOT framework~\cite{brun1997,antcheva2009}, and processes the partonic ROOT $n$-tuples generated directly by the code described in Sec.~\ref{sec:event_generation}.

\begin{figure}
  \centering
  \begin{subfigure}{0.49\textwidth}
    \centering
    \includegraphics[width=\textwidth]{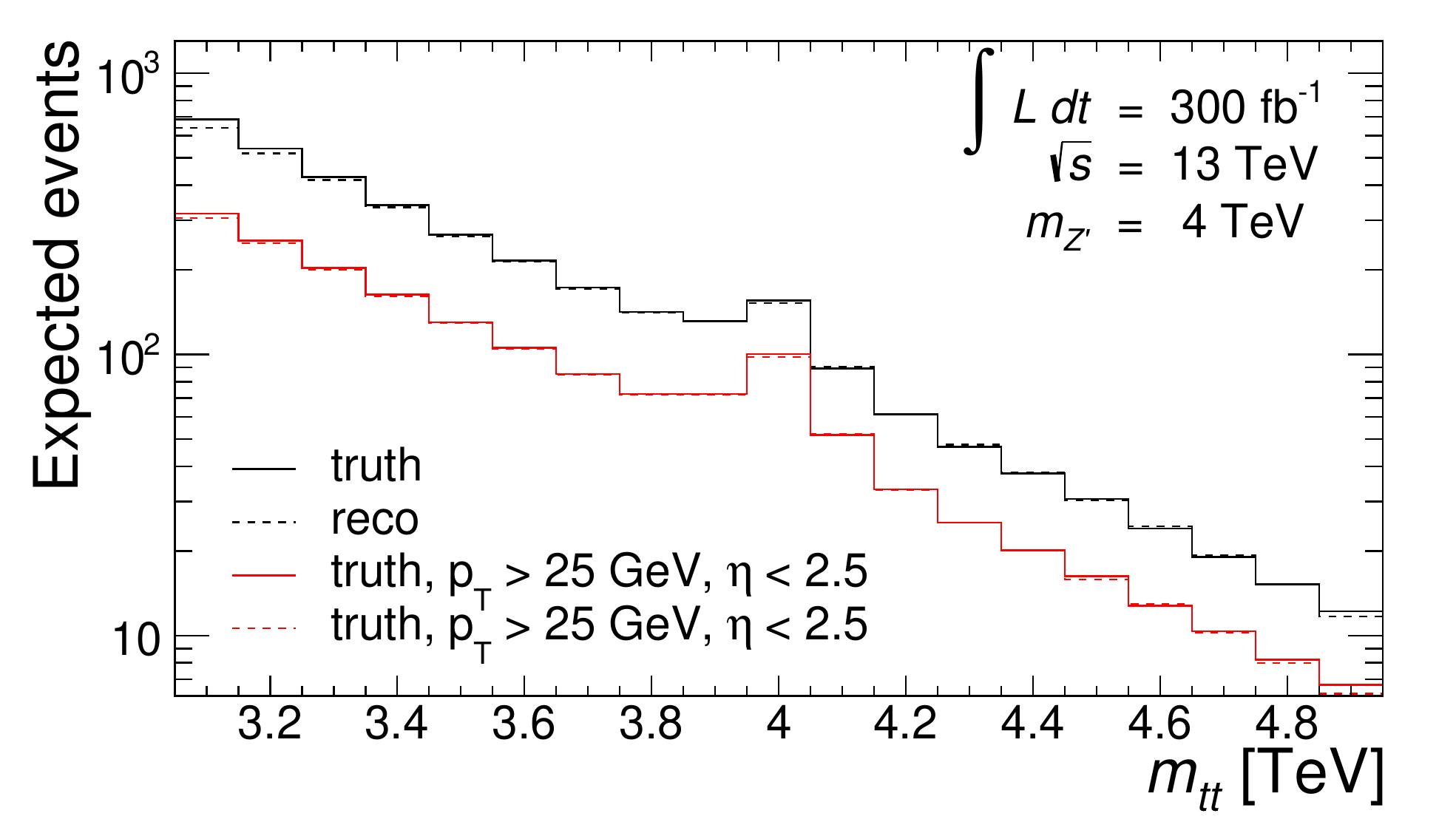}
    \caption{Expected events}
    \vspace{0.5cm}
  \end{subfigure}
  \begin{subfigure}{0.49\textwidth}
    \centering
    \includegraphics[width=\textwidth]{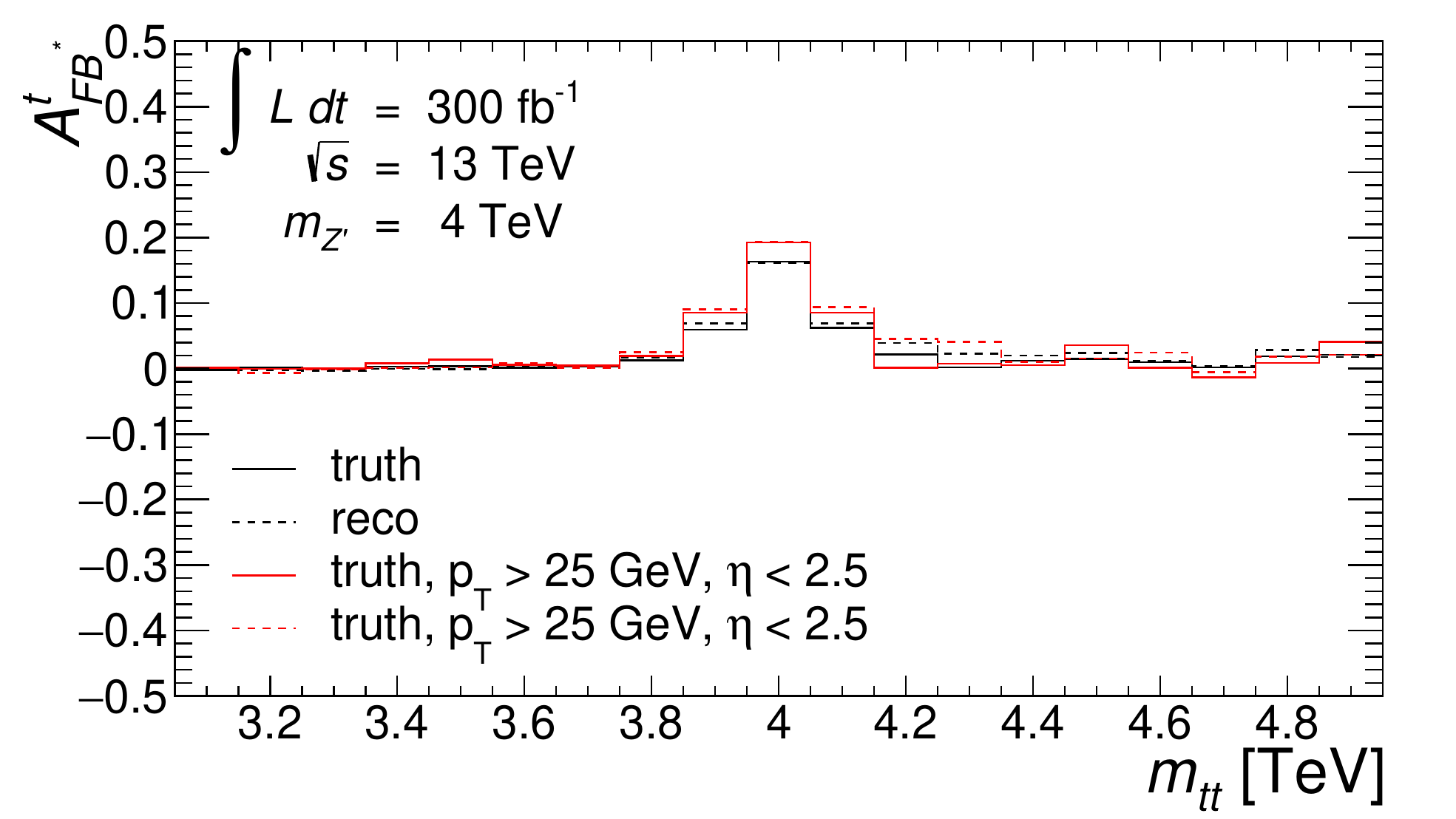}
    \caption{$A^{t}_{FB^{*}}$}
    \vspace{0.5cm}
  \end{subfigure}
  \begin{subfigure}{0.49\textwidth}
    \centering
    \includegraphics[width=\textwidth]{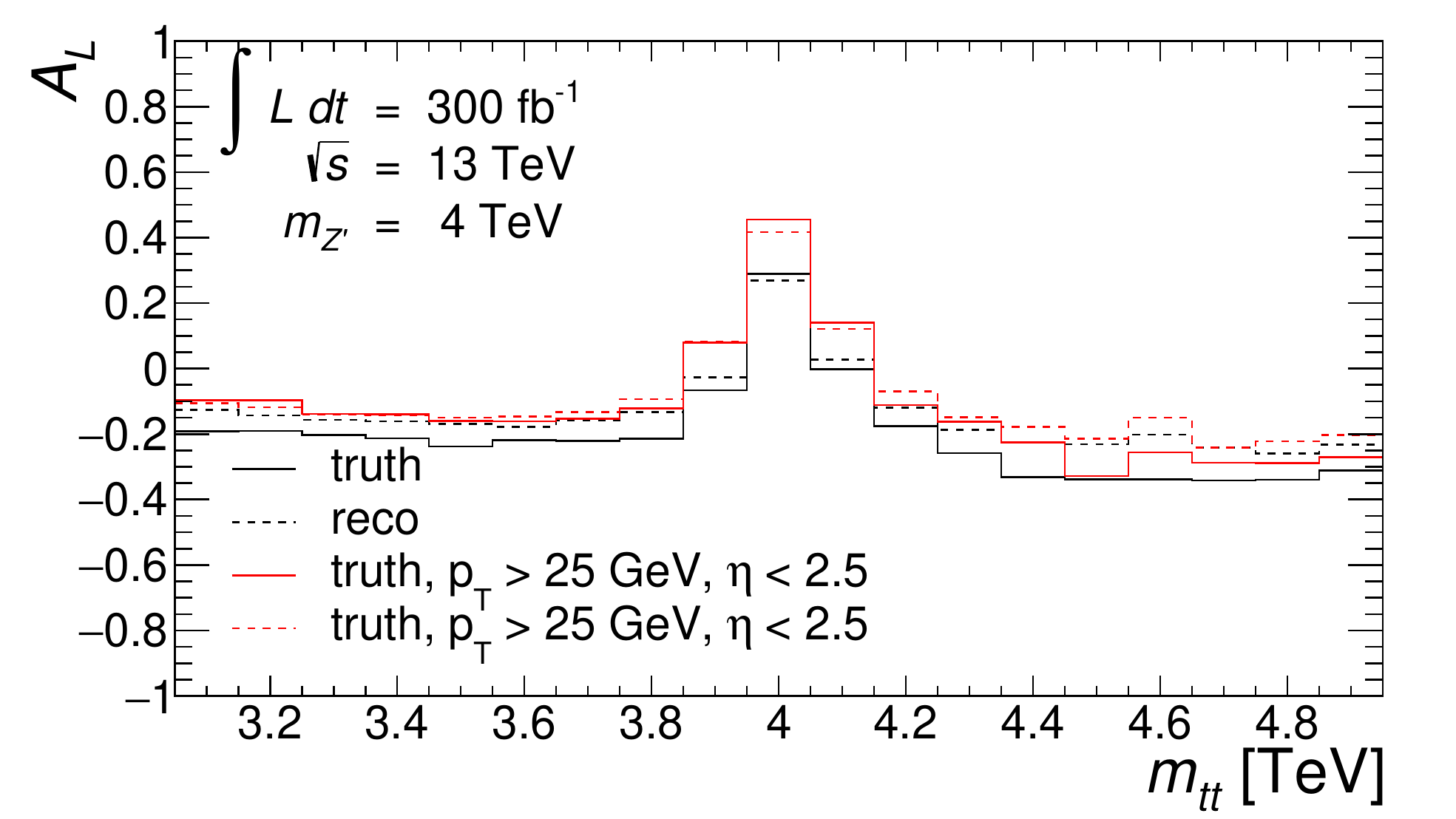}
    \caption{$A_{L}$}
  \end{subfigure}
  \caption{Plots comparing the results of the toy reconstruction (Sec.~\ref{sec:top_reconstruction}) with the truth for the cross-section, $A^{t}_{FB^{*}}$, and $A_L$, expected with an integrated luminosity of $300$~fb$^{-1}$, at $\sqrt{s}=13$~TeV, with and without fiducial cuts. The example model is GLR-R, with the pole mass of the $Z'$ fixed at $4$~TeV.}
  \label{fig:top_reconstruction}
\end{figure}

\begin{figure}
  \centering
  \begin{subfigure}{0.49\textwidth}
    \includegraphics[width=\textwidth]{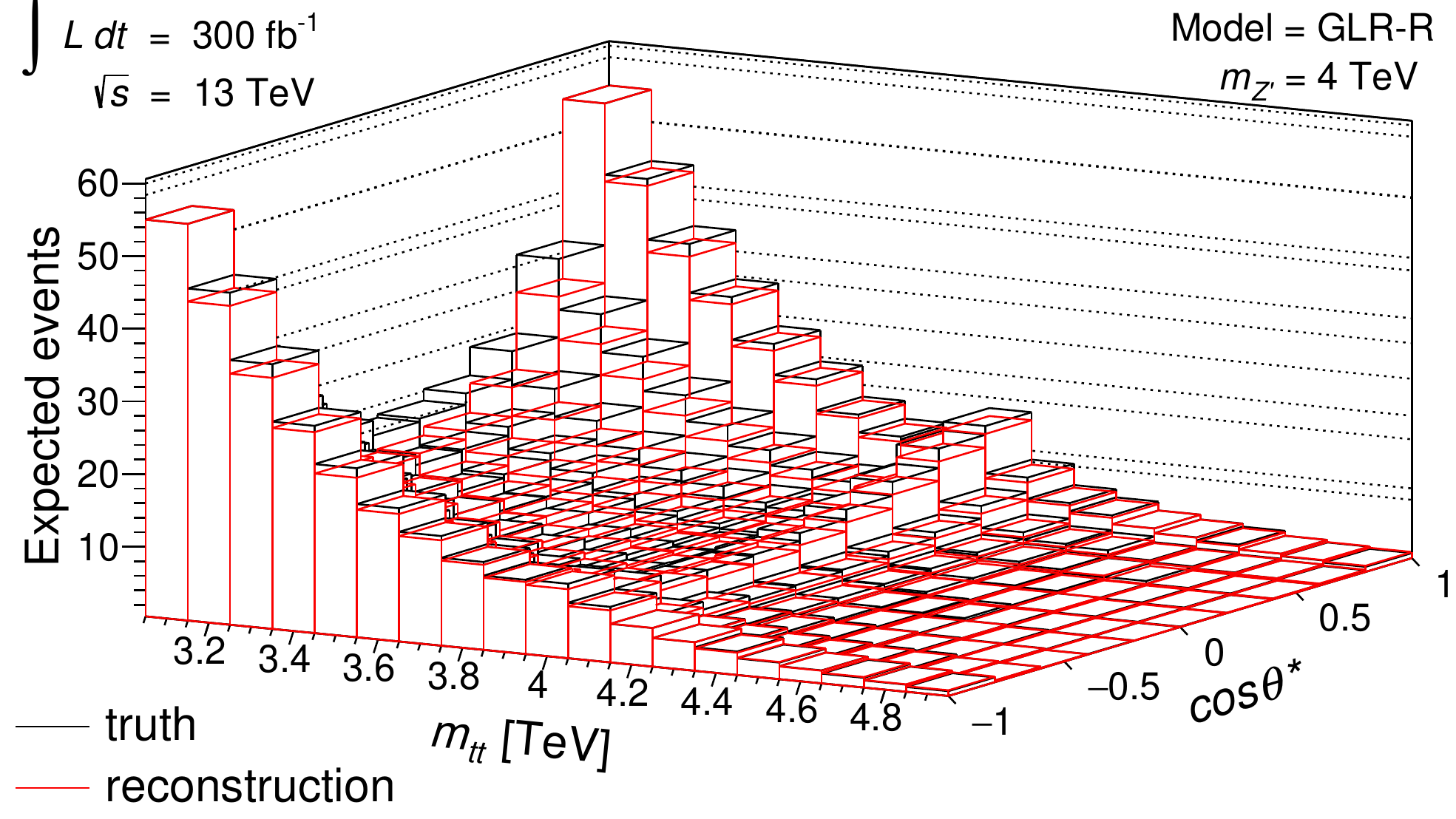}
    \caption{$\cos\theta^{*}$}
  \end{subfigure}
  \begin{subfigure}{0.49\textwidth}
    \includegraphics[width=\textwidth]{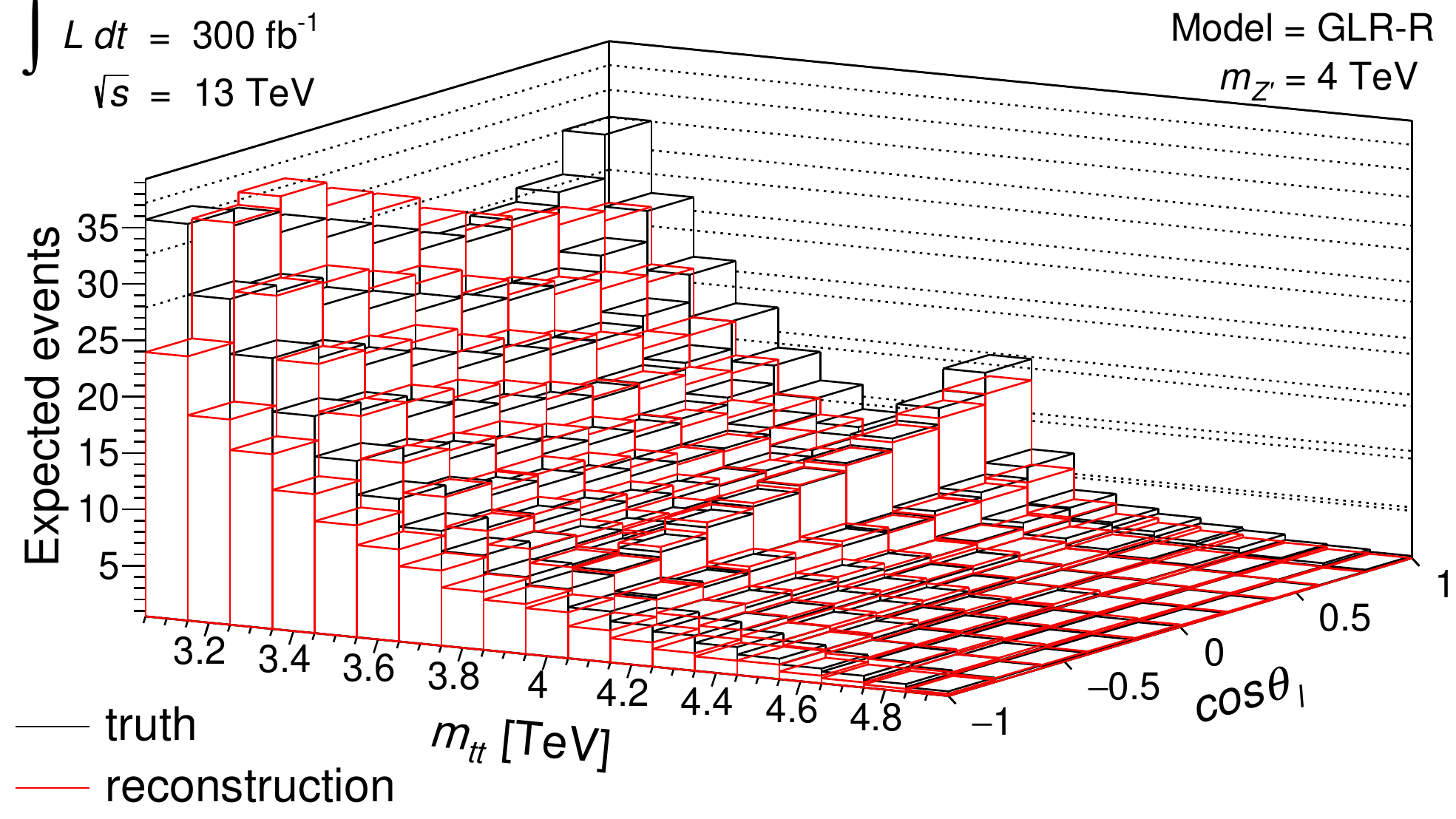}
    \caption{$\cos\theta_{\ell}$}
  \end{subfigure}
  \caption{Plots comparing the results of the toy reconstruction (Sec.~\ref{sec:top_reconstruction}) with the truth for the cross-section, $A^{t}_{FB^{*}}$, and $A_L$, expected with an integrated luminosity of $300$~fb$^{-1}$, at $\sqrt{s}=13$~TeV. The example model is GLR-R, with the pole mass of the $Z'$ fixed at $4$~TeV.}
  \label{fig:top_reconstruction_2d}
\end{figure}

A simplified kinematic requirement on every final state particle of at least $25$ GeV $p_{T}$, and $|\eta| < 2.5$ is applied. The results of this toy reconstruction, with and without fiducial cuts, for the three observables of interest ($\sigma$, $A^{t}_{FB^*}$, and $A_{L}$) are presented in Fig.~\ref{fig:top_reconstruction}. They show that, even accounting for the experimental constraints described above, each of the observables retains its characteristic shape. However, it should be reiterated that this scenario is an optimistic one, disregarding detector efficiencies, and neglecting additional jets from initial/final state radiation, which will significantly increase the complexity of the reconstruction. All subsequent plots show only variables that have undergone this toy reconstruction. Fig.~\ref{fig:top_reconstruction_2d} shows that the toy reconstruction skews somewhat the 2D distributions: biasing events away from zero in $\cos\theta^{*}$ and towards zero for $\cos\theta_{\ell}$.

As the $Z'$ signal arises only from quark-antiquark fusion, where the phase space favours $q$ having a higher parton momentum fraction than $\bar{q}$, it has been common to include a requirement on the rapidity of the top pair: $|y_{tt}| > 0.5$~\cite{basso2012a,baumgart2013}. This enhances the new physics signal over the SM production by excluding events with a dominant contribution from gluon-gluon interactions. However, it also reduces the number of signal events, resulting in a drop in significance that somewhat counters the effect. For example, the significance for the U$(1)_{R}$ model drops from $8.1\sigma$ to $~4\sigma$ for a 1D analysis in $m_{tt}$, and from $8.7\sigma$ to $~4.5\sigma$ for a 2D analysis with $\cos\theta$, when including a $0.5$ cut on $|y_{tt}|$. Consequently, for the resonant masses and luminosities considered in this study, no requirement is made on $y_{tt}$.

\subsection{Expected significance}
\label{sec:expected_significance}

In order to characterise the sensitivity of an LHC experiment to each of these $Z'$ models, the significance that a benchmark data analysis would achieve is calculated, assuming these models describe Nature. Models that hypothesise a greater observational deviation from the SM generally predict a higher significance and, consequently, better motivate an LHC search. For our purposes, the null hypothesis ($H_0$) includes only the known $t\bar{t}$ processes of the SM. The alternative hypotheses ($H$) include the SM processes with the addition of a single $Z'$ for each BSM scenario. Therefore, the signal cross section ($\sigma_s$) is defined:
\begin{equation}
  \sigma_{s} = \sigma_{(Z'+t\bar{t})} - \sigma_{t\bar{t}} = \sigma_{Z'} + \sigma_{int(Z',t\bar{t})},
\end{equation}
where $\sigma_{t\bar{t}}$ denotes the cross section for $t\bar{t}$ production in the SM only, and $\sigma_{Z'}$ as the $t\bar{t}$ cross section when mediated solely by a $Z'$, with $\sigma_{int(Z',t\bar{t})}$ as the cross section corresponding exclusively to the interference term. The signal, therefore, comprises the isolated $Z'$ contribution and the interference with the SM.

For each simulated event a number of kinematic variables (${\bf x}$) may be chosen. Constructing a histogram, $\textbf{n}=\{n_i\}$, in one or more of these variables, the expectation value for each bin ($i$) in signal ($s_i$) and background ($b_i$) contributions may be linearly separated as
\begin{equation}
  \nu_{i} = \mu s_i + b_i,
  \label{eq:nu_i}
\end{equation}
where the $\mu$ parameter determines the strength of the signal process, with mean frequency
\begin{equation}
  s_i(s_{tot},\boldsymbol{\theta}_s) = s_{tot}\alpha_i(\boldsymbol{\theta}_s) = s_{tot} \int_{x_i^{min}}^{x_i^{max}} f_s (\mathbf{x};\boldsymbol{\theta}_s)\,d\mathbf{x}.
  \label{eq:s_i}
\end{equation}
Here $\alpha_i$ represents the probability of finding an event with $\mathbf{x}$ in bin $i$, while $s_{tot}$ is the total mean number of signal events. The probability density function (pdf) for $\mathbf{x}$ is denoted by $f_s$, where $\theta_s$ encodes the parameters that dictate the shape of $f_s$. An analogous relation exists also for $b_i$. $\boldsymbol{\theta}$ denotes all the parameters determining the shape of $\textbf{n}$.

A suitable statistic for a test of $\mu = 0$ must be constructed, under the assumption of $\mu = 1$, i.e. where the simulated data for the BSM is playing the part of the experimental data, while the simulated SM data provides the hypothesis under investigation. Rejecting $\mu=0$ signifies the discovery of a BSM signal, and the suitability of this model to motivate an LHC search. In order to test for the presence of new physics frequentist inference is used: the level of agreement of the observed data with a given hypothesis is quantified by computing the $p$-value, using the likelihood ratio as a test statistic. The likelihood function is constructed as the product of a Poisson distribution for all bins:
\begin{equation}
  L(\mathbf{x} | \mu, \boldsymbol{\theta}) = \prod_{i}e^{\nu_i}\frac{\nu_i^{n_i}}{n_i!},
\end{equation}
from which the profile likelihood ratio is
\begin{equation}
  \lambda(\mu) = \frac{L(\mu, \hat{\hat{\boldsymbol{\theta}}})}{L(\hat{\mu}, \hat{\boldsymbol{\theta}})}.
  \label{eq:1d_likelihood_ratio}
\end{equation}
Here $\hat{\hat{\boldsymbol{\theta}}}$ represents the $\mu$ dependent maximum likelihood (ML) estimator of $\boldsymbol{\theta}$, while $\hat{\mu}$ and $\hat{\boldsymbol{\theta}}$ are the unconditional estimators of their respective variables. In order to test $\mu = 0$ it is convenient to define
\begin{equation}
  q_0 =
  \begin{cases}
    -2\ln\lambda(0) & \hat{\mu} \ge 0, \\
    0               & \hat{\mu} < 0.
  \end{cases}
\end{equation}
In defining the statistic above, it is implicitly assumed that $\mu \ge 0$, i.e. the signal can only increase the mean frequency above what would be expected from $H_0$. Notice that higher values of $q_{0}$ imply increasing disagreement between the observed data (represented by the alternate hypothesis $H$) and $H_0$. From $q_0$ a measure of the agreement between $H_0$ and $H$ may be directly quantified by recovering the $p$-value as
\begin{equation}
  p_0 = \int^\infty_{q_{0,obs}} f(q_0|0)\,dq_0.
\end{equation}
Therefore, in order to determine $p_0$, the sampling distribution $f(q_0|0)$ is required. An approximation for the profile Likelihood ratio may be found in the large sample limit, as described in~\cite{cowan2011}. These asymptotic formulae allow the significance for $H$, and the full sampling distribution, to be determined without involving computationally expensive Monte Carlo calculations.

As is common practice, the $p$-value is converted into the equivalent ``$\sigma$ value''; assuming the distribution of the test statistic under $H_0$ follows a normal distribution, this is the number ($Z$) of standard deviations ($\sigma$) the result lies above the mean such that its upper-tail probability is equal to $p_0$,
\begin{equation}
  Z = \Phi^{-1}(1-p_0),
\end{equation}
where $\Phi$ is the cumulative normal distribution. In the collider physics community, a significance of $5\sigma$ is generally considered necessary for rejecting $H_0$ and hailing a discovery of new physics.

The foremost kinematic variable used in a search for new physics is the invariant mass of the final state system $\sqrt{\hat{s}}$, which for a $t\bar{t}$ process equates to the mass of the top pair ($m_{t\bar{t}}$). One-dimensional (1D) event distributions, binned in $m_{tt}$, are the dominant discovery tool in any search for new resonances. Due to the cross section dependence on $\sqrt{\hat{s}}$ (Eq.~\ref{eq:Mqqtt2}), this ``cut and count'' methodology is known as ``bump-hunting.'' This 1D distribution is given for all benchmark models, along with the corresponding significance.

In addition to the usual bump-hunt, this study expands the analysis to include a number of asymmetry observables, as described in Secs.~\ref{sec:charge_asymmetry} and~\ref{sec:top_polarisation_asymmetry}. The dominant use of these observables is to profile a newly discovered $Z'$ by evaluating which models best describe their shape. However, the potential for these asymmetries to act as complementary discovery observables to the invariant mass distribution is also explored here.

This is done by binning events in two-dimensional (2D) histograms, where the second variable corresponds to the defining observable of the asymmetry, i.e. for $A^{t}_{FB^{*}}$, both $m_{tt}$ and $\cos\theta^*$ are used, while for $A_{L}$, bins are made in $m_{tt}$ and $\cos\theta_{\ell}$. Note that Eqs.~\ref{eq:nu_i} and \ref{eq:s_i} are completely general, hence, the mean frequency in a given 2D bin ($i,j$) is simply
\begin{equation}{}
  \nu_{i,j}(\mu, \sigma_{t\bar{t}{}},\sigma_{Z'},\boldsymbol{\theta}) = L_{int}[\epsilon_{i,j}(\theta_b)\sigma_{t\bar{t}} + \mu\alpha_{i,j}(\theta_s)(\sigma_{Z'}+\sigma_{int\{Z',t\bar{t}\}})],
\end{equation}

with a corresponding likelihood function. Here $L_{int}$ represents the integrated luminosity, $\epsilon$ and $\alpha$ represent the efficiencies for the SM background and the signal, respectively, to fall in the bin $i,j$. This is dependent on both the asymmetry introduced by the model in question, and the efficiencies of the detector.

Methods to perform the expected significance evaluation, as described above are available in RooStats~\cite{moneta2010}, with the formalism and numerical implementation of these methods described in~\cite{cowan2011}. Tools to generate the statistical models for RooStats from the generated histograms were provided by HistFactory~\cite{cranmer2012}.

\section{Results}
\label{sec:results}

A selection of results are presented that profile the benchmark $Z'$ models, using the observables described in the previous section. These are the cross section ($\sigma$), in the form of the expected number of events, the reconstructed forward-backward asymmetry ($A^{t}_{FB^{*}}$), and the top quark polarisation asymmetry ($A_L$). Each of these variables are binned as a function of $m_{tt}$. Considering the current limits on the mass of the $Z'$ (Sec.~\ref{sec:summary_and_present_limits}) the $Z'$ mass is fixed at $4$~TeV. The centre of mass energy of the LHC is simulated at $13$~TeV. All results assume the collection of a total integrated luminosity ($L_{int}$) of $300$~fb$^{-1}$. The statistical error is quantified for this luminosity by binning the expected number of events ($L\sigma$) in $m_{tt}$ for a bin width of $100$~GeV, and assuming Poisson errors, i.e. $\delta N = \sqrt{N} = \sqrt{\sigma L_{int}}$. The $A_{FB}$ observable, as the division of subtracted/summed events classified in to two exclusive subsets $N_1$ and $N_2$, under the assumption that the two are independent, has a statistical uncertainty
\begin{equation}
  \delta A = \frac{2}{N}\sqrt{\frac{N_1 N_2}{N}} = \sqrt{\frac{1 - A^2}{\sigma L_{int}}}.
\end{equation}
While it is expected that the dominant source of uncertainty is statistical, it is likely that there will be also be significant systematic uncertainties. These may be addressed when this study is extended to include the full parton-shower, hadronisation and detector reconstruction, in addition to the required efficiencies, and the assessment of the significance provides this functionality.

In the following, the distributions for the differential cross section, $A^{t}_{FB^{*}}$ and $A_{L}$, for the benchmark $Z'$ models are presented, highlighting their power to distinguish different classes of model from the SM and each other. Following this the 1D significances and 2D significances are given for each model stressing the scope of asymmetry observables to act as complementary discovery observables for specific models.

\subsection{Distinguishing \texorpdfstring{$Z'$}{Z'} using asymmetries}

\begin{figure}
  \centering
  \begin{subfigure}{0.49\textwidth}
    \includegraphics[width=\textwidth]{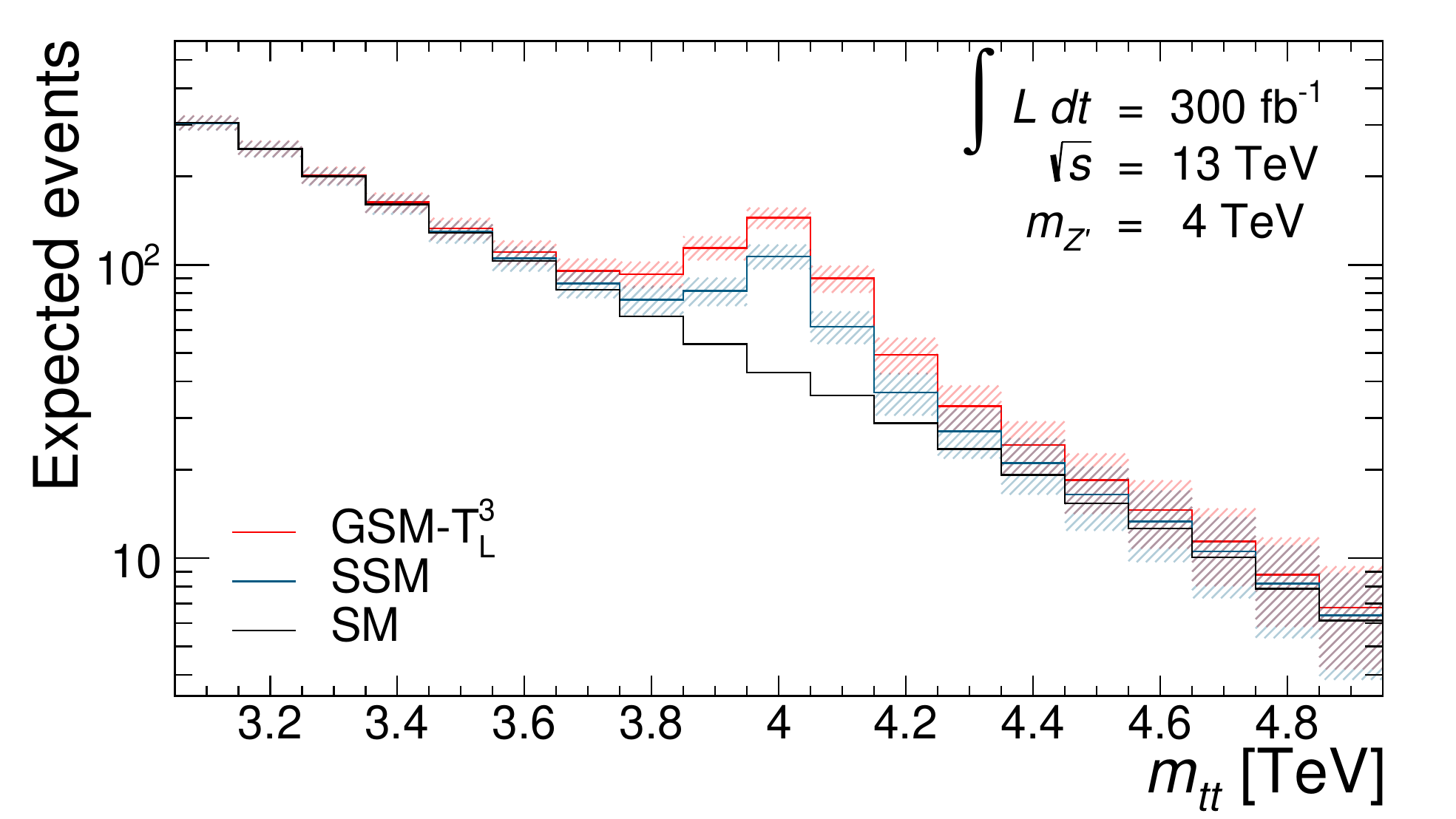}
    \caption{Events expected -- GSM models}
    \vspace{1cm}
  \end{subfigure}
  \begin{subfigure}{0.49\textwidth}
    \includegraphics[width=\textwidth]{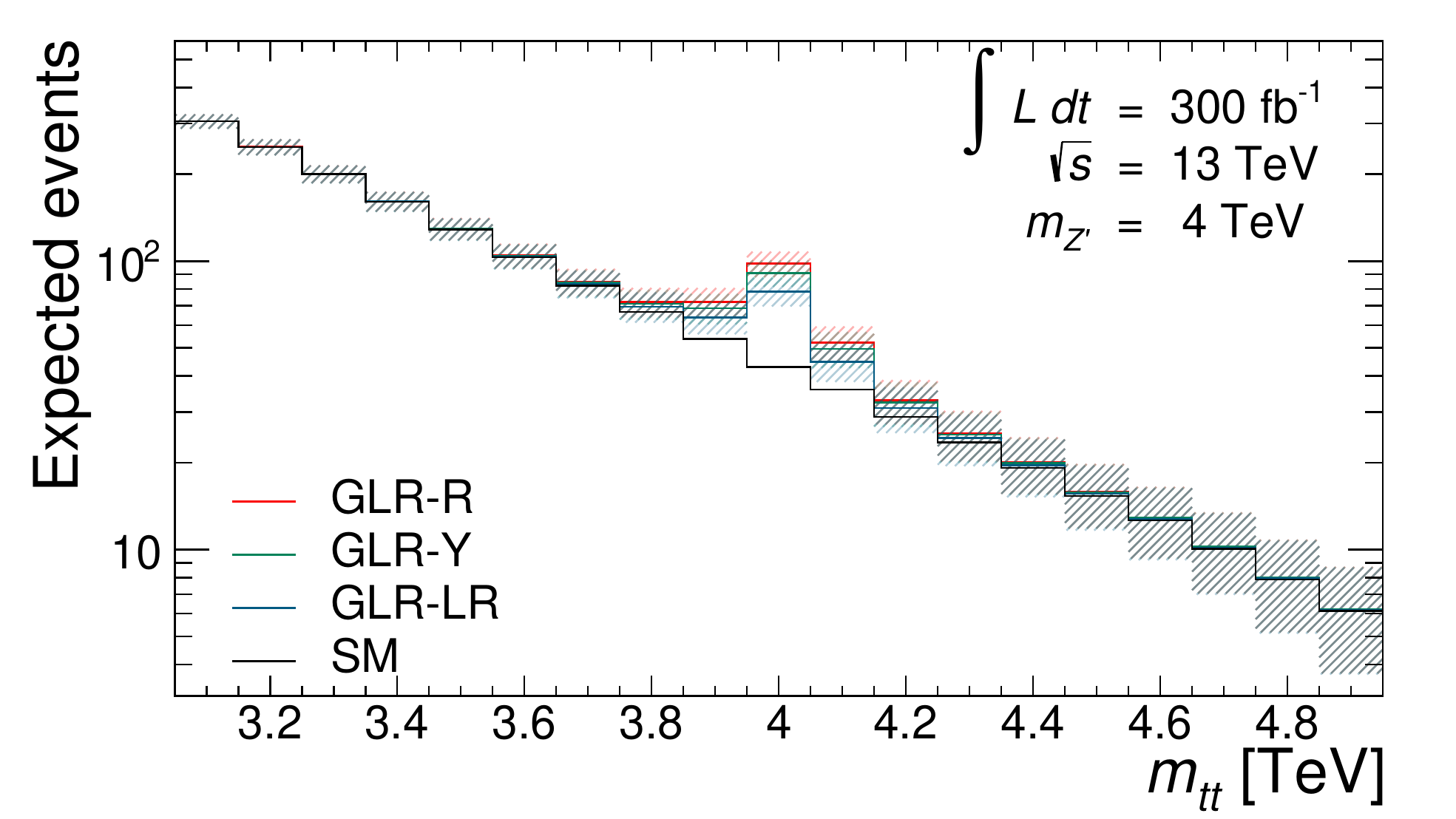}
    \caption{Events expected -- GLR models}
    \vspace{1cm}
  \end{subfigure}
  \begin{subfigure}{0.49\textwidth}
    \includegraphics[width=\textwidth]{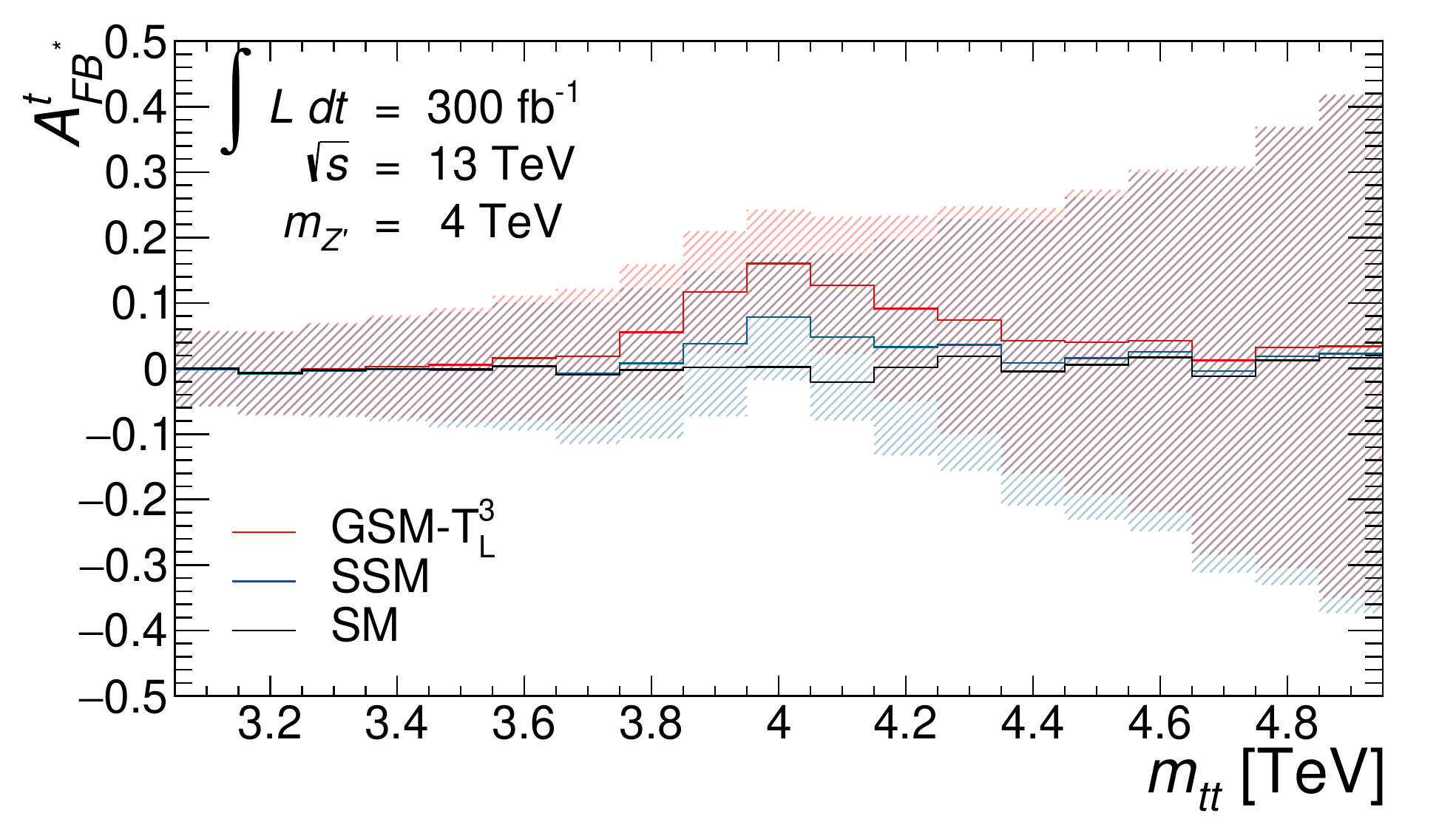}
    \caption{$A^{t}_{FB^{*}}$ -- GSM models}
    \vspace{1cm}
  \end{subfigure}
  \begin{subfigure}{0.49\textwidth}
    \includegraphics[width=\textwidth]{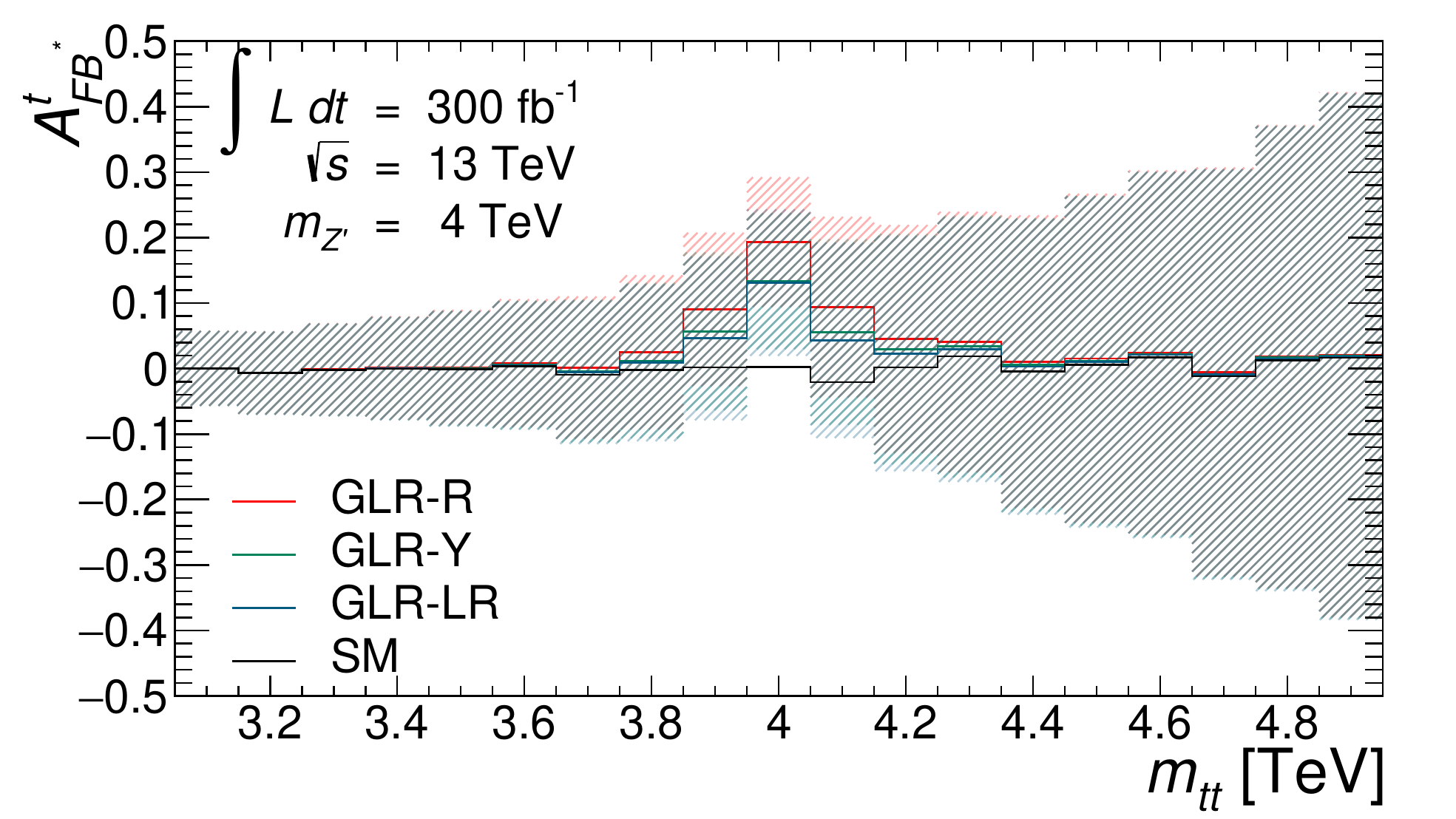}
    \caption{$A^{t}_{FB^{*}}$ -- GLR models}
    \vspace{1cm}
  \end{subfigure}
  \begin{subfigure}{0.49\textwidth}
    \includegraphics[width=\textwidth]{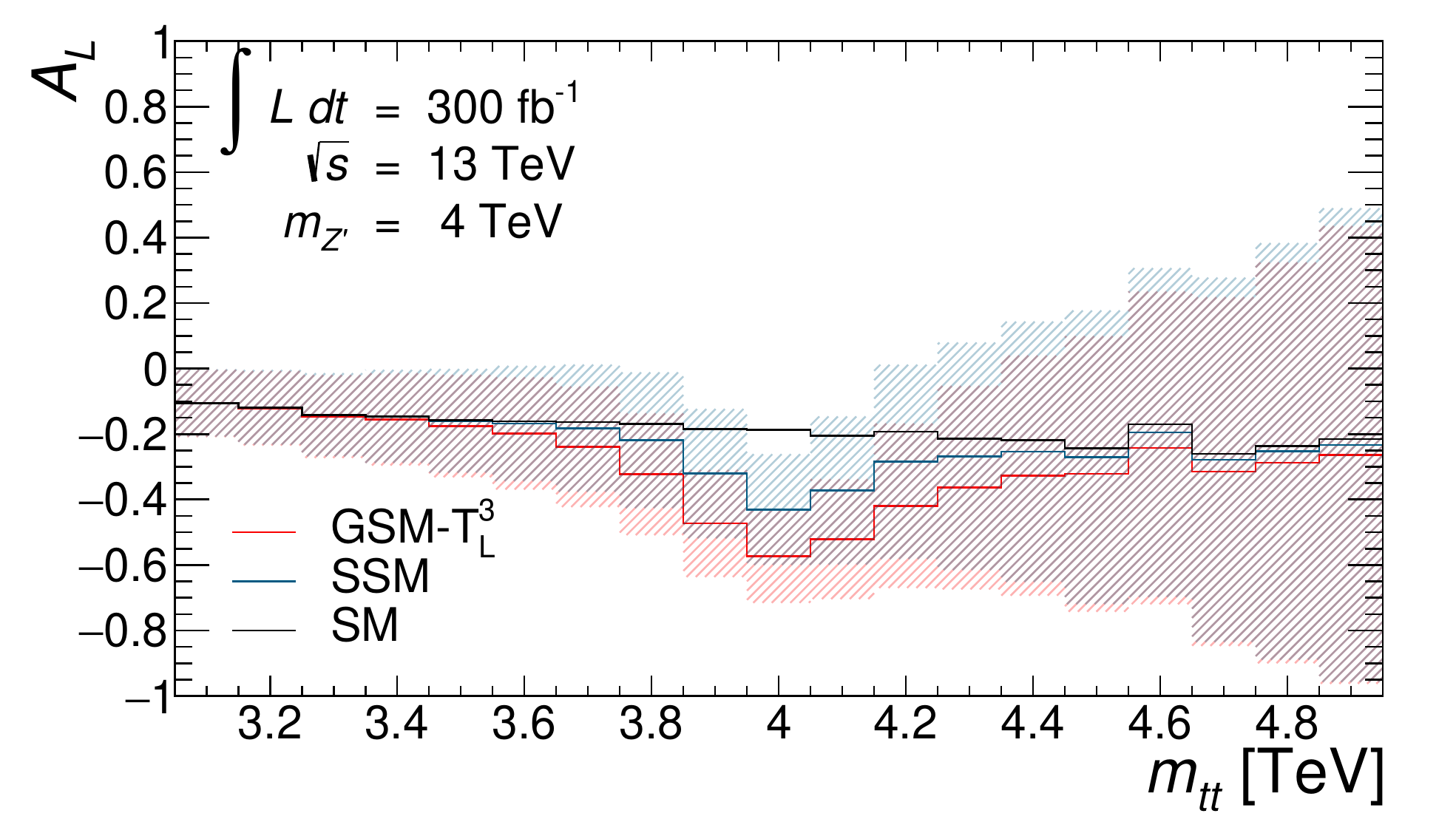}
    \caption{$A_L$ -- GSM models}
    \vspace{1cm}
  \end{subfigure}
  \begin{subfigure}{0.49\textwidth}
    \includegraphics[width=\textwidth]{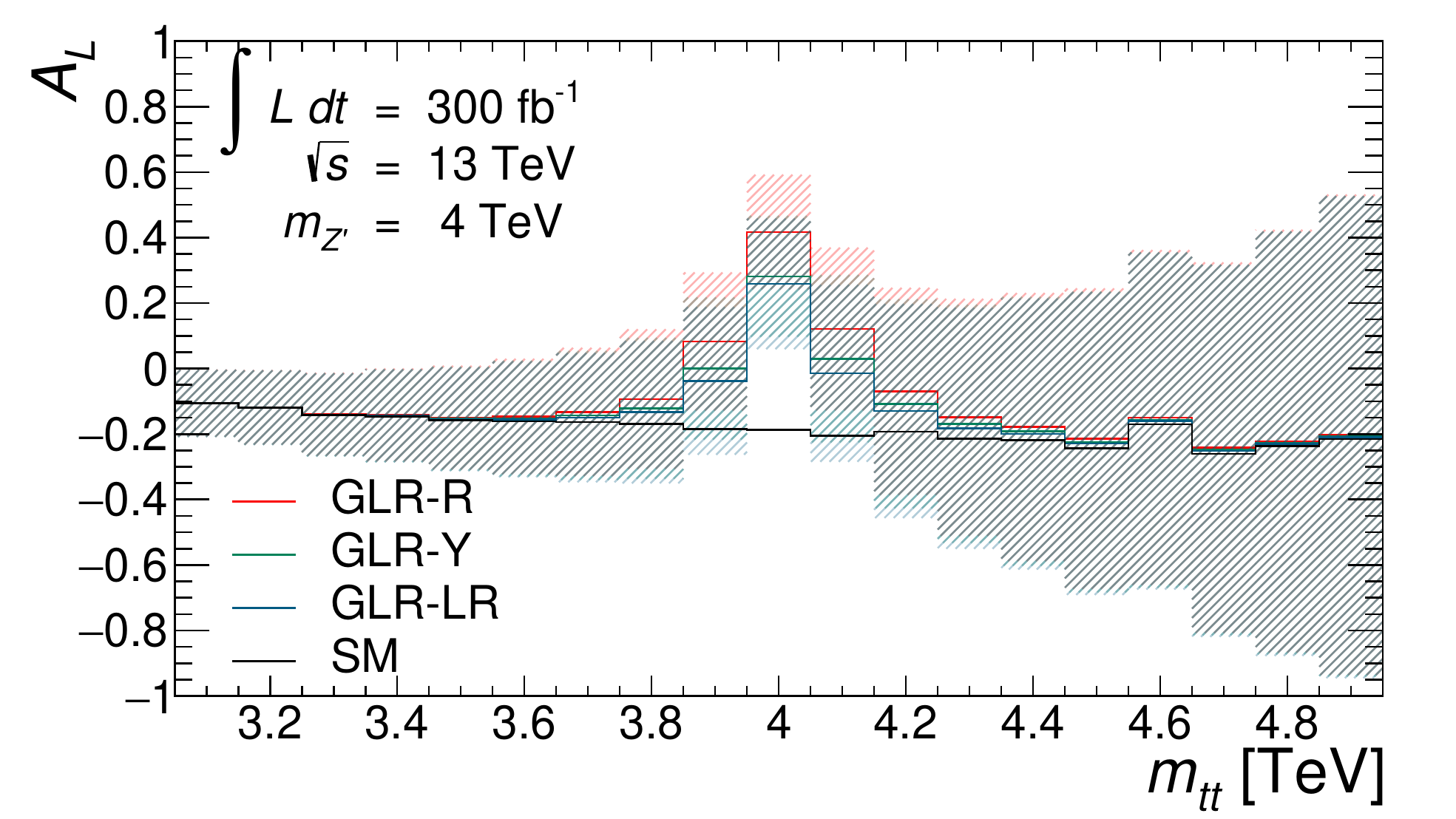}
    \caption{$A_L$ -- GLR models}
    \vspace{1cm}
  \end{subfigure}
  \caption{Expected distributions for each of our observables of interest, with an integrated luminosity of $300$~fb$^{-1}$, at $\sqrt{s}=13$~TeV. The pole mass of the $Z'$ is fixed at $4$~TeV for every model. The shaded bands indicate the projected statistical uncertainty, assuming Poisson errors on the number of events (Sec.~\ref{sec:results}). Of our benchmark model set, only this subset has non-trivial asymmetries. Shown from top to bottom are: the events expected (cross section) [a,b]; $A^{t}_{FB^{*}}$ [c,d]; and $A_{L}$ [e,f], all as function of $m_{tt}$. GSM-type models are shown on the left [a,c,e]; and GLR-type models on the right [b,d,f].}
  \label{fig:distinguishing}
\end{figure}

\begin{figure}
  \centering
  \begin{subfigure}{0.49\textwidth}
    \includegraphics[width=\textwidth]{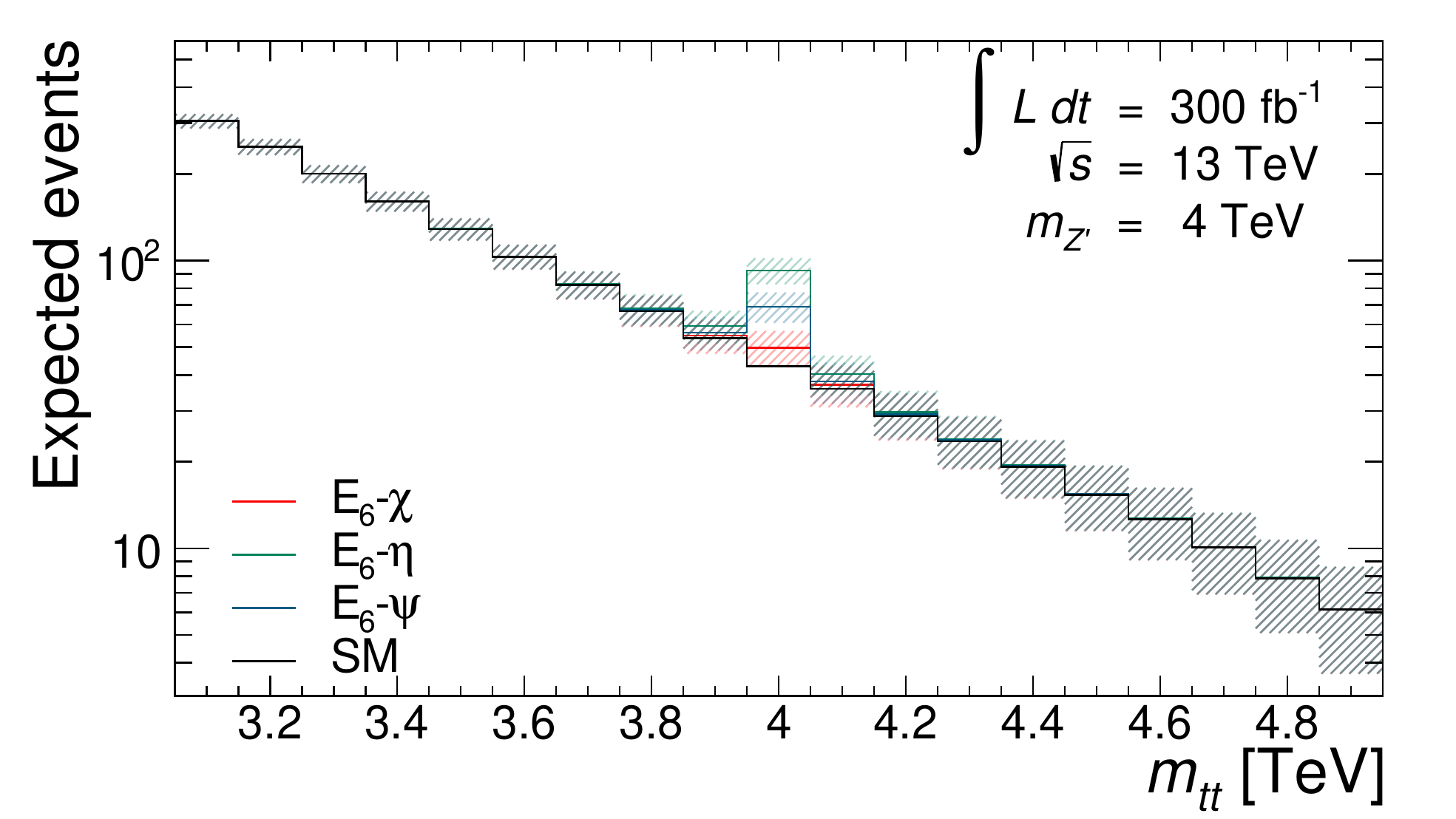}
    \caption{Expected events -- E$_6$ models (I)}
    \vspace{0.5cm}
  \end{subfigure}
   \begin{subfigure}{0.49\textwidth}
    \includegraphics[width=\textwidth]{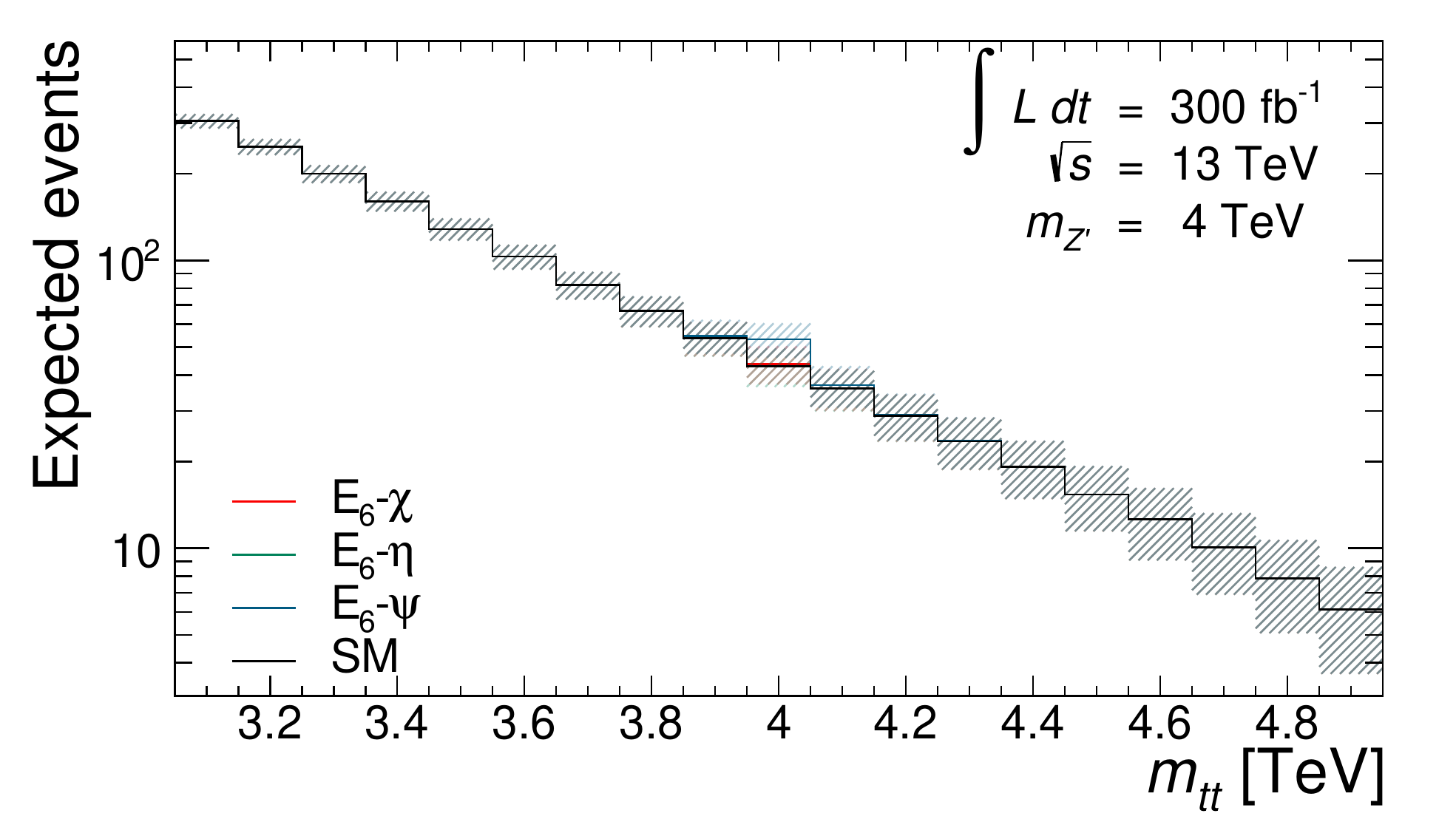}
    \caption{Expected events -- E$_6$ models (II)}
    \vspace{0.5cm}
  \end{subfigure}
  \begin{subfigure}{0.49\textwidth}
    \includegraphics[width=\textwidth]{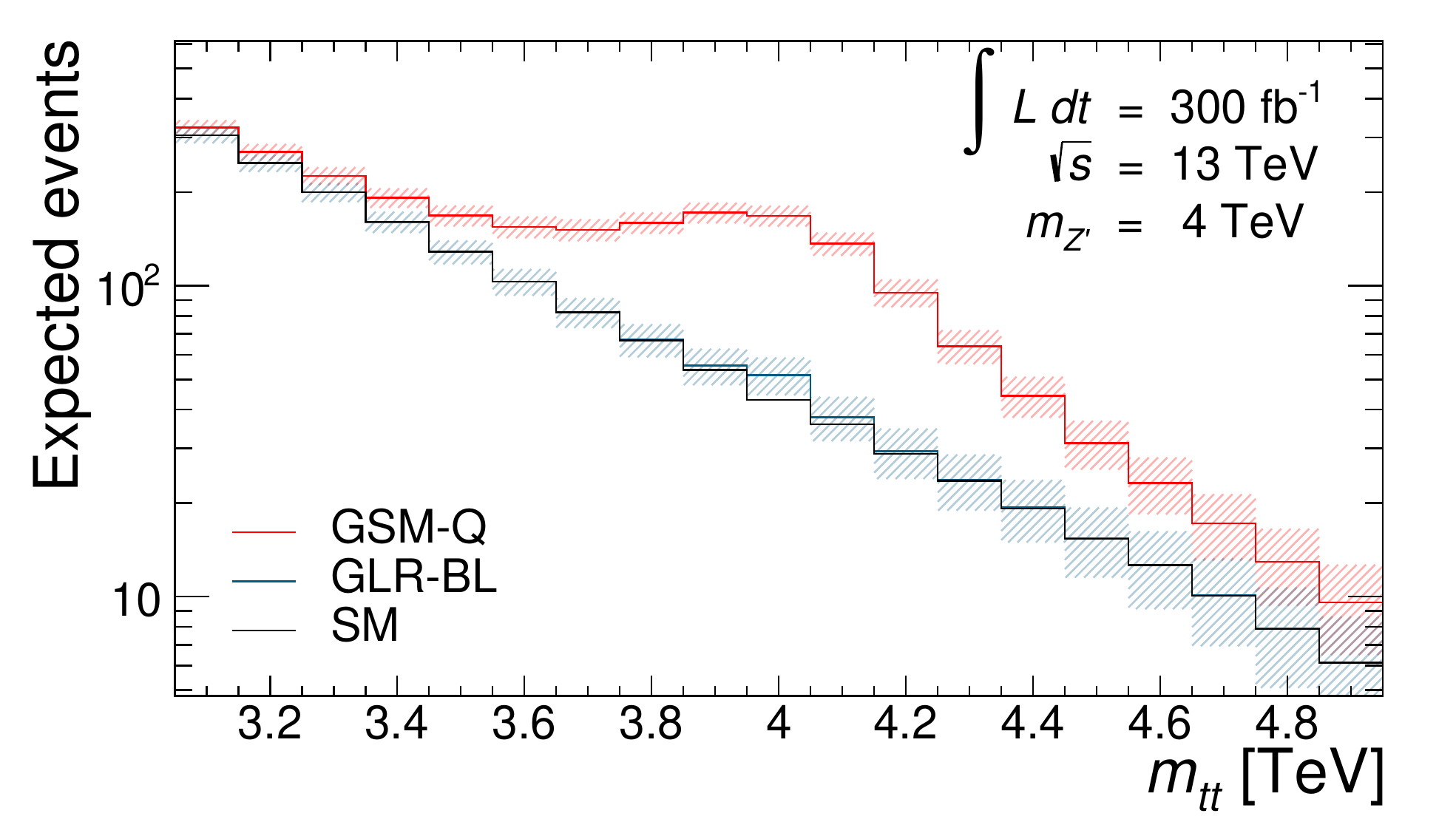}
    \caption{Expected events --  GSM-Q \& GLR-BL models}
  \end{subfigure}
  \caption{Expected $m_{tt}$ distributions for models with negligible $A^{t}_{FB^{*}}$ and $A_{L}$, with an integrated luminosity of $300$~fb$^{-1}$, at $\sqrt{s}=13$~TeV. The pole mass of the $Z'$ is fixed at $4$~TeV for every model. The shaded bands indicate the projected statistical uncertainty, assuming Poisson errors on the number of events (Sec.~\ref{sec:results}).}
  \label{fig:distinguishing2}
\end{figure}

Figure~\ref{fig:distinguishing} shows plots for the expected events (differential cross section), $A^{t}_{FB^{*}}$ and $A_{L}$ as a function of $m_{tt}$, for two models from the GSM class, and three from the GLR class. These models feature no vanishing axial and vector couplings to top quarks (Tab.~\ref{tab:zprime_parameters}), and, therefore, result in notable $A_{L}$ and $A^{t}_{FB^{*}}$ (Eq.~\ref{eq:afb_coupling}, \ref{eq:al_coupling}). The remaining models, including all of the $E_6$ class, only produce an asymmetry via the interference term, which generally gives an undetectable enhancement with respect to the SM yield. Consequently, these distributions are not shown. However, the expected number of events, binned in $m_{tt}$, for the remaining models is shown in Fig.~\ref{fig:distinguishing2}.

The cross section, profiled in $m_{tt}$, shows a very visible peak for all models in Figs.~\ref{fig:distinguishing}a and \ref{fig:distinguishing}b. The GSM models feature a greater peak, and width, consistent with their stronger couplings, but the impact on the cross section is otherwise similar for both classes. Mirroring the cross section, the shape of the $A^{t}_{FB^{*}}$ distribution clearly distinguishes the models and the SM, with the difference in width even more readily apparent (Figs.~\ref{fig:distinguishing}c and \ref{fig:distinguishing}d). The best distinguishing power over all the models investigated comes from the $A_{L}$ distribution, which features an oppositely signed peak for the GLR and GSM classes (Figs.~\ref{fig:distinguishing}e and \ref{fig:distinguishing}f).

The heavy photon scenario defined by the GSM-Q model in Fig~\ref{fig:distinguishing2} shows a very strong response, in fact, a highly significant mass peak coupled with no response in $A^{t}_{FB^{*}}$ or $A_{L}$ would be a strong signature for a model of this type. The model featuring a pure U$(1)_{B-L}$ symmetry, results in a negligible peak, so is of little phenomenological interest. The $E_6$ class of models universally features a zero vector coupling to up type quarks and, therefore, has negligible asymmetry response. Additionally, the $S$, $I$ and $N$ realisations feature only a very small increase in the number of events on peak, while the $\chi$, $\eta$ and $\phi$ scenarios result in a narrow resonance. The absence of a corresponding peak in either asymmetry would also be supportive of these models, offering an additional handle on diagnosing a discovered $Z'$.

\subsection{Asymmetries as complementary discovery variables}

Using the statistical techniques outlined in Sec.~\ref{sec:expected_significance}, the significance for each of these models are given assuming them to exist in Nature, against a test of the SM null hypothesis. These methods are applied to the 1D histograms, binned as a function of $m_{tt}$, as presented in Figs.~\ref{fig:distinguishing}a, \ref{fig:distinguishing}b and \ref{fig:distinguishing2}. This gives an assessment of the distinguishing power of the standard bump-hunt for each of these models, for comparison with the 2D analysis, and to show the relative potential for observation in experiment.

\begin{figure}
  \centering
  \begin{subfigure}{0.49\textwidth}
    \includegraphics[width=\textwidth]{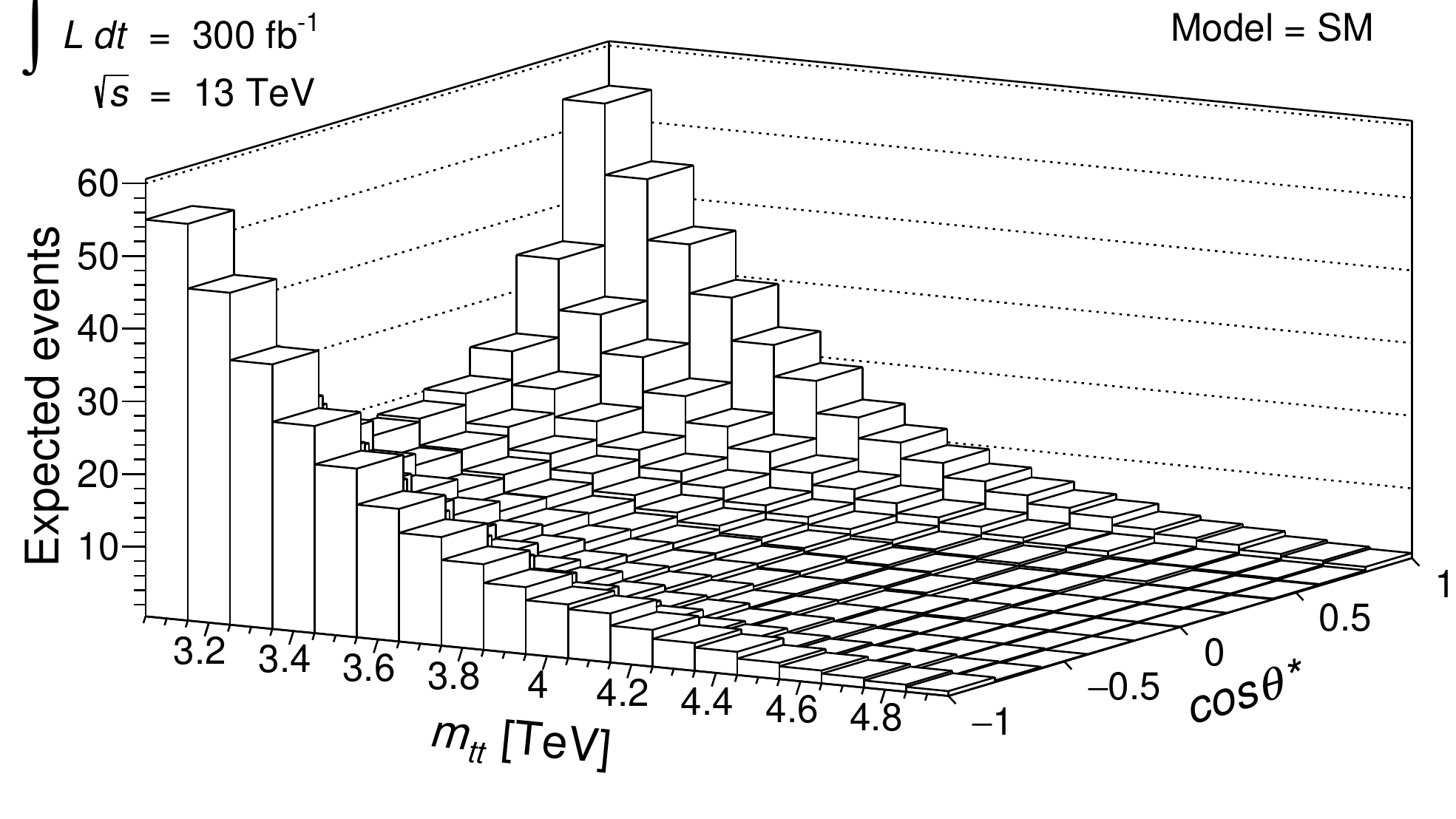}
    \caption{SM}
    \vspace{1cm}
  \end{subfigure}
  \begin{subfigure}{0.49\textwidth}
    \includegraphics[width=\textwidth]{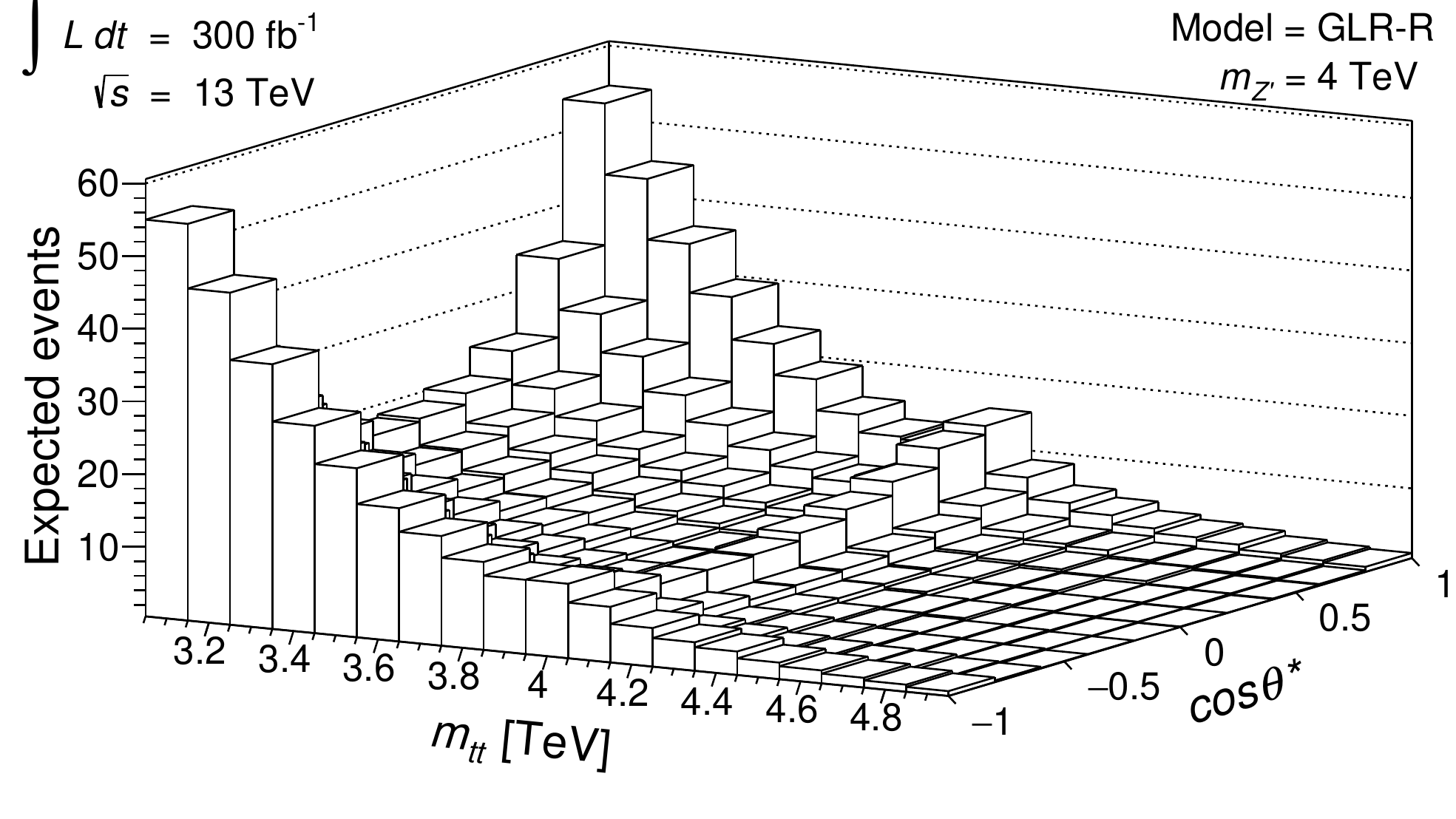}
    \caption{GLR-R}
    \vspace{1cm}
  \end{subfigure}
  \begin{subfigure}{0.49\textwidth}
    \includegraphics[width=\textwidth]{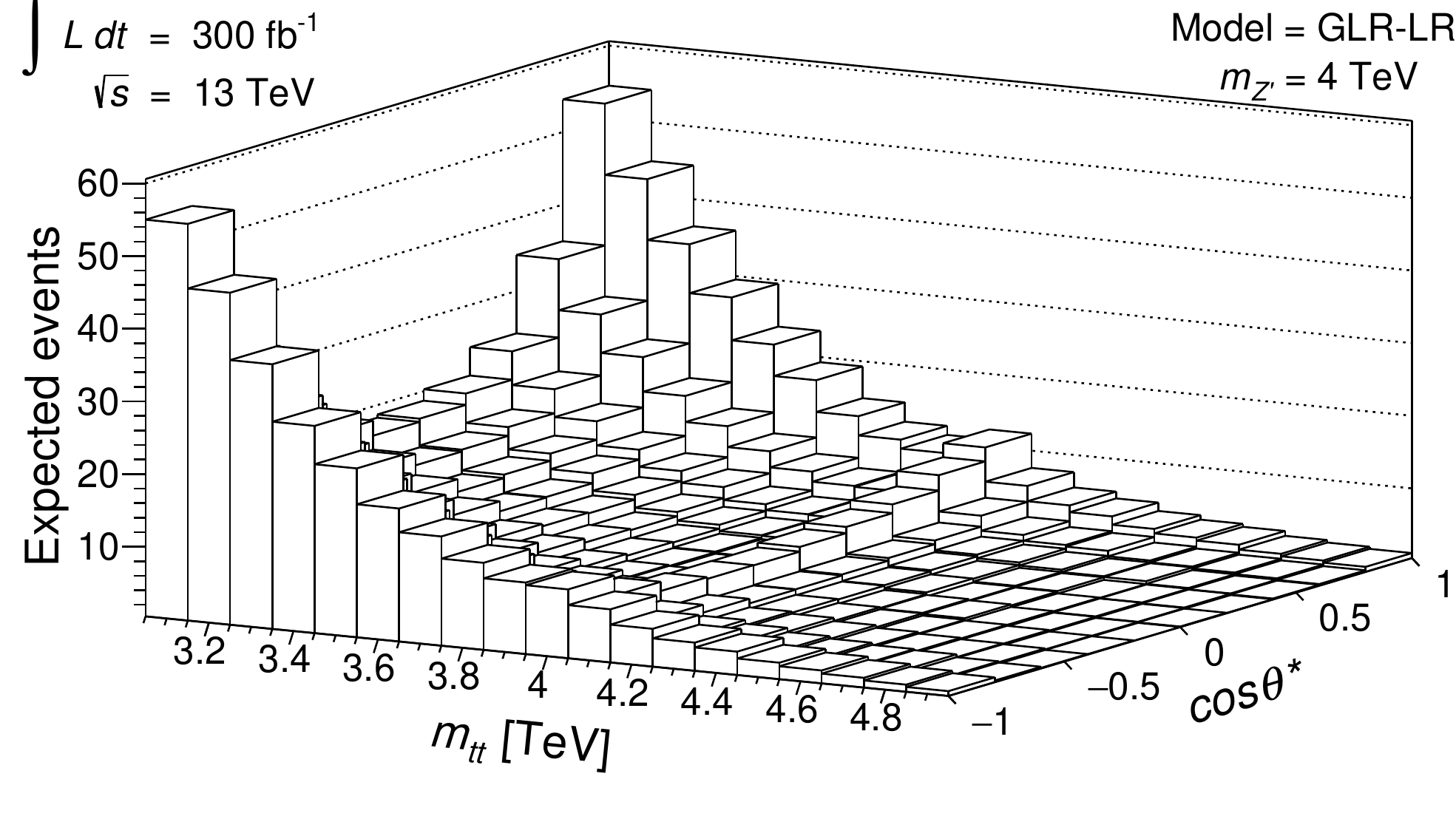}
    \caption{GLR-LR}
    \vspace{1cm}
  \end{subfigure}
  \begin{subfigure}{0.49\textwidth}
    \includegraphics[width=\textwidth]{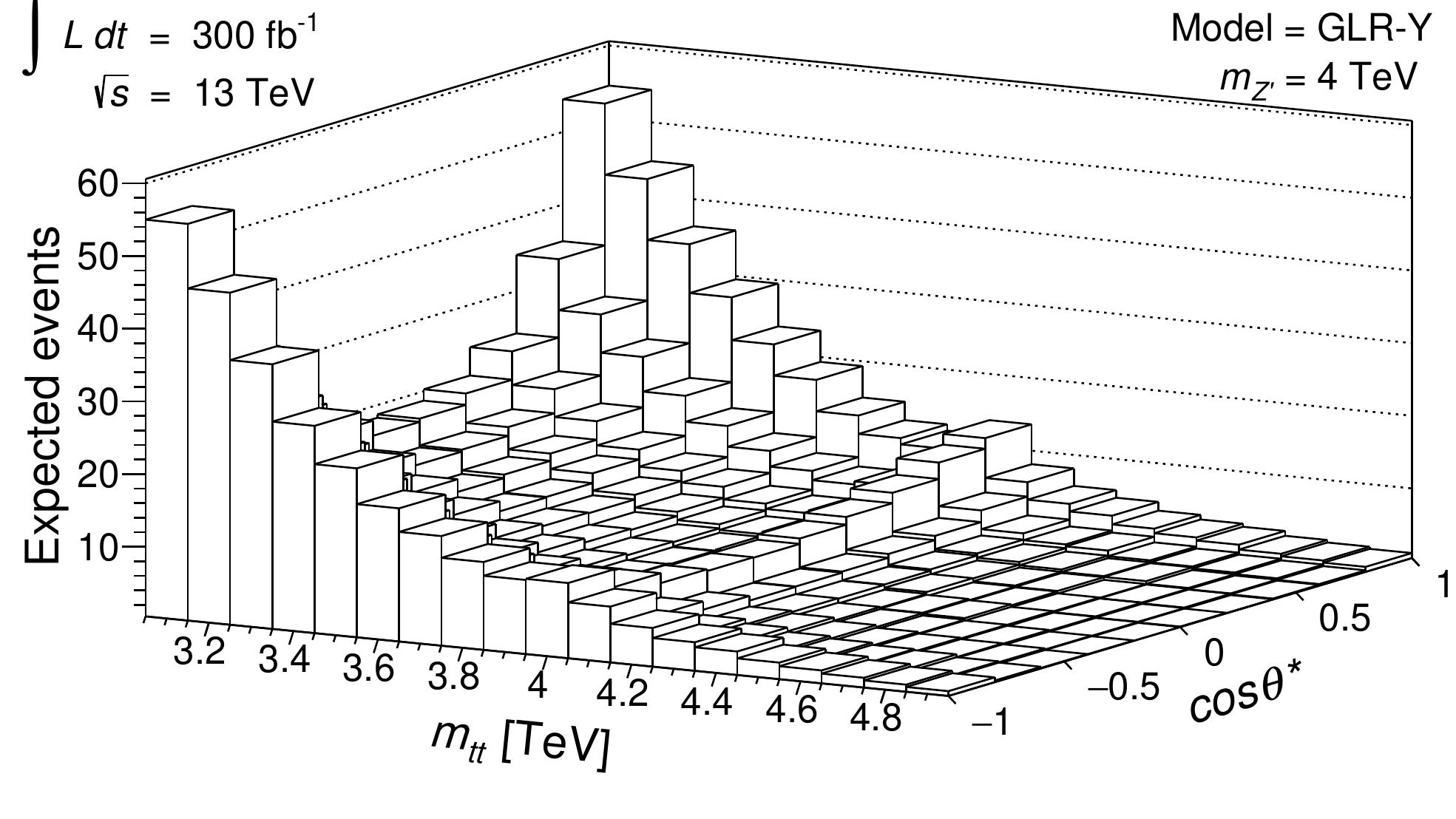}
    \caption{GLR-Y}
    \vspace{1cm}
  \end{subfigure}
  \begin{subfigure}{0.49\textwidth}
    \includegraphics[width=\textwidth]{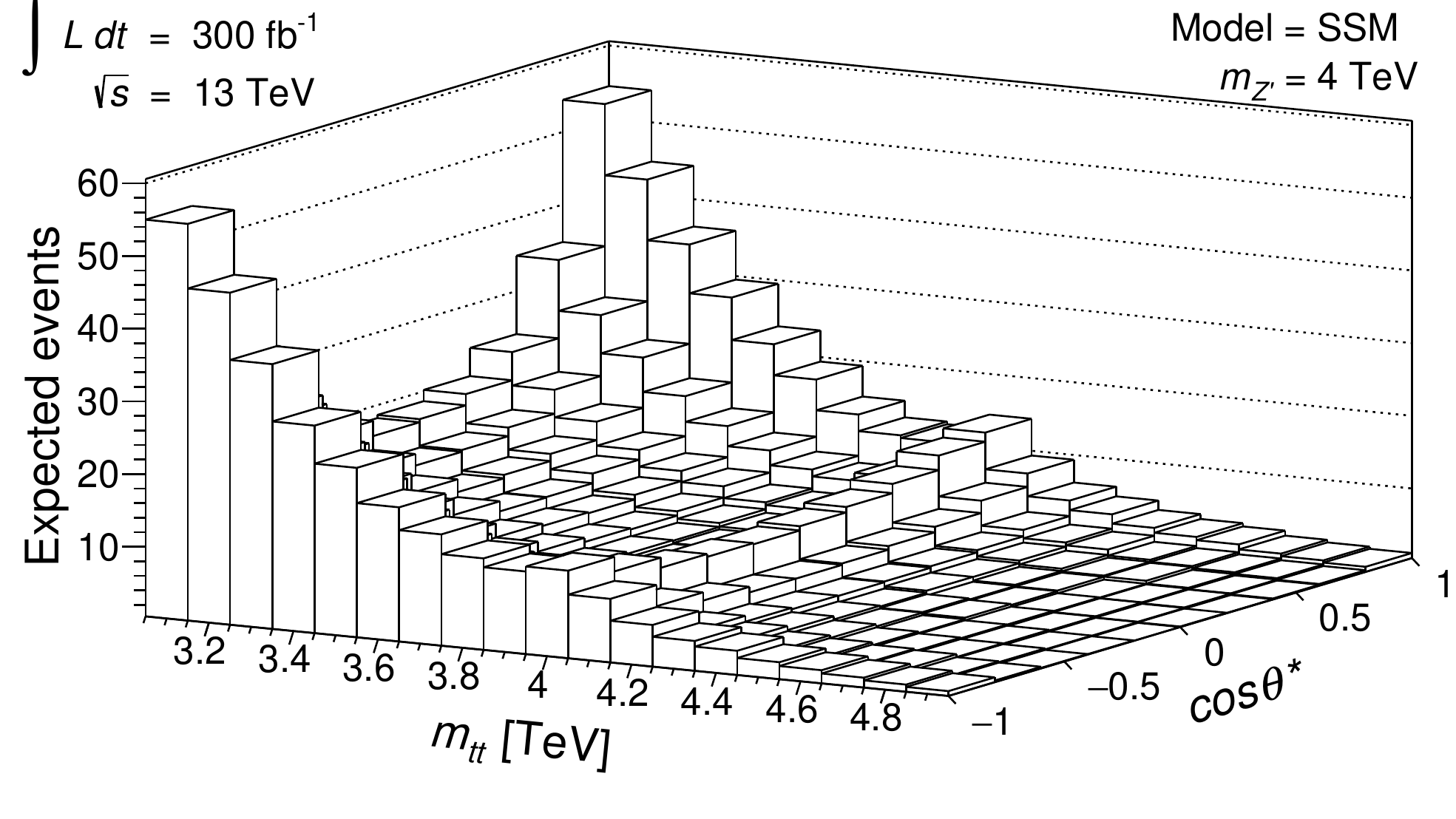}
    \caption{SSM}
    \vspace{1cm}
  \end{subfigure}
  \begin{subfigure}{0.49\textwidth}
    \includegraphics[width=\textwidth]{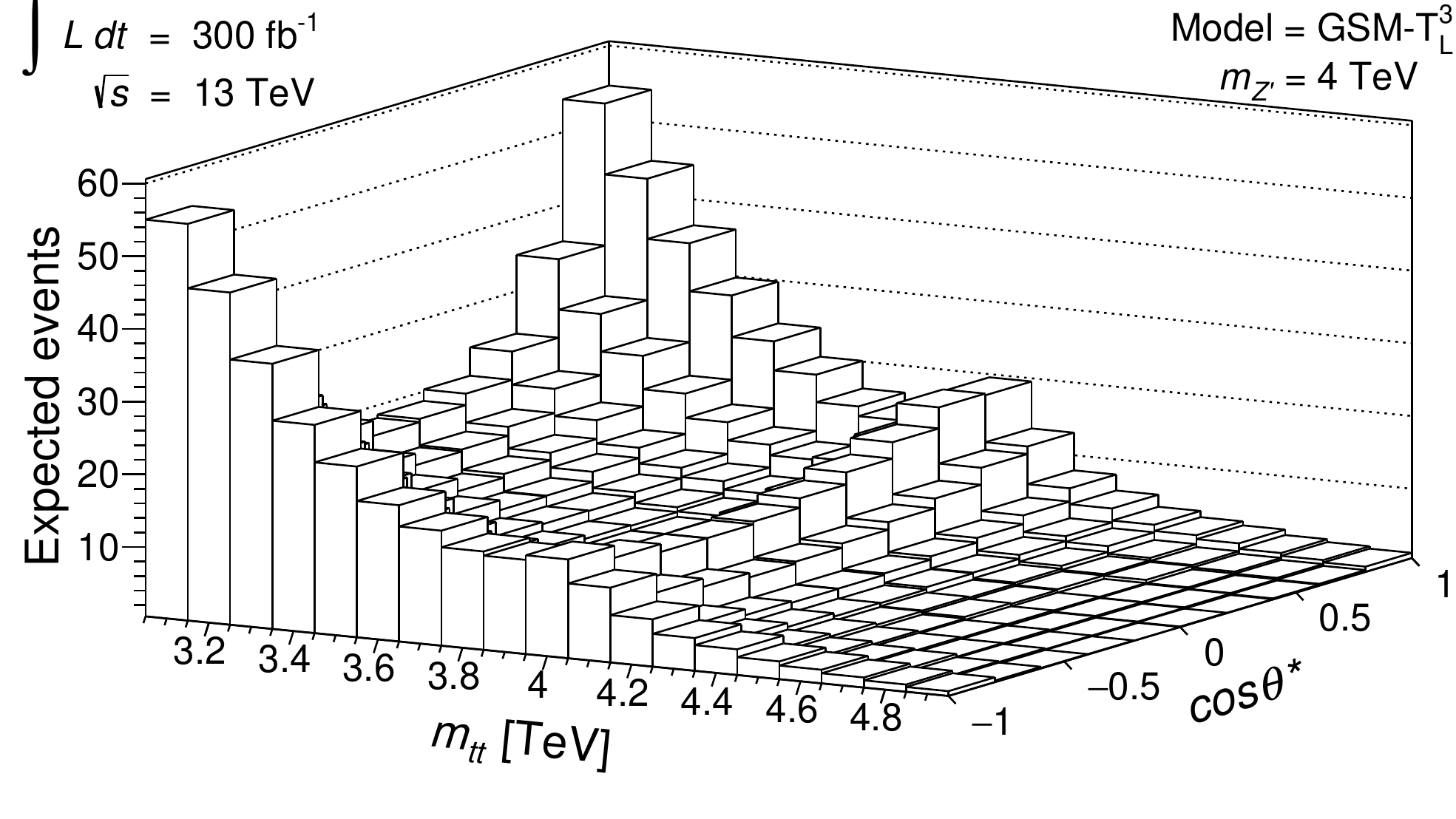}
    \caption{GSM-$T^3_L$}
    \vspace{1cm}
  \end{subfigure}
  \caption{Expected number of events binned in both $m_{tt}$ and $\cos\theta^*$ after toy reconstruction. These plots are used to derive the 2D significances in these variables, and access the combined significance of using the $m_{tt}$ distribution and $A^{t}_{FB^{*}}$.}
  \label{fig:mtt_costhetastar}
\end{figure}

\begin{figure}
  \centering
  \begin{subfigure}{0.49\textwidth}
    \includegraphics[width=\textwidth]{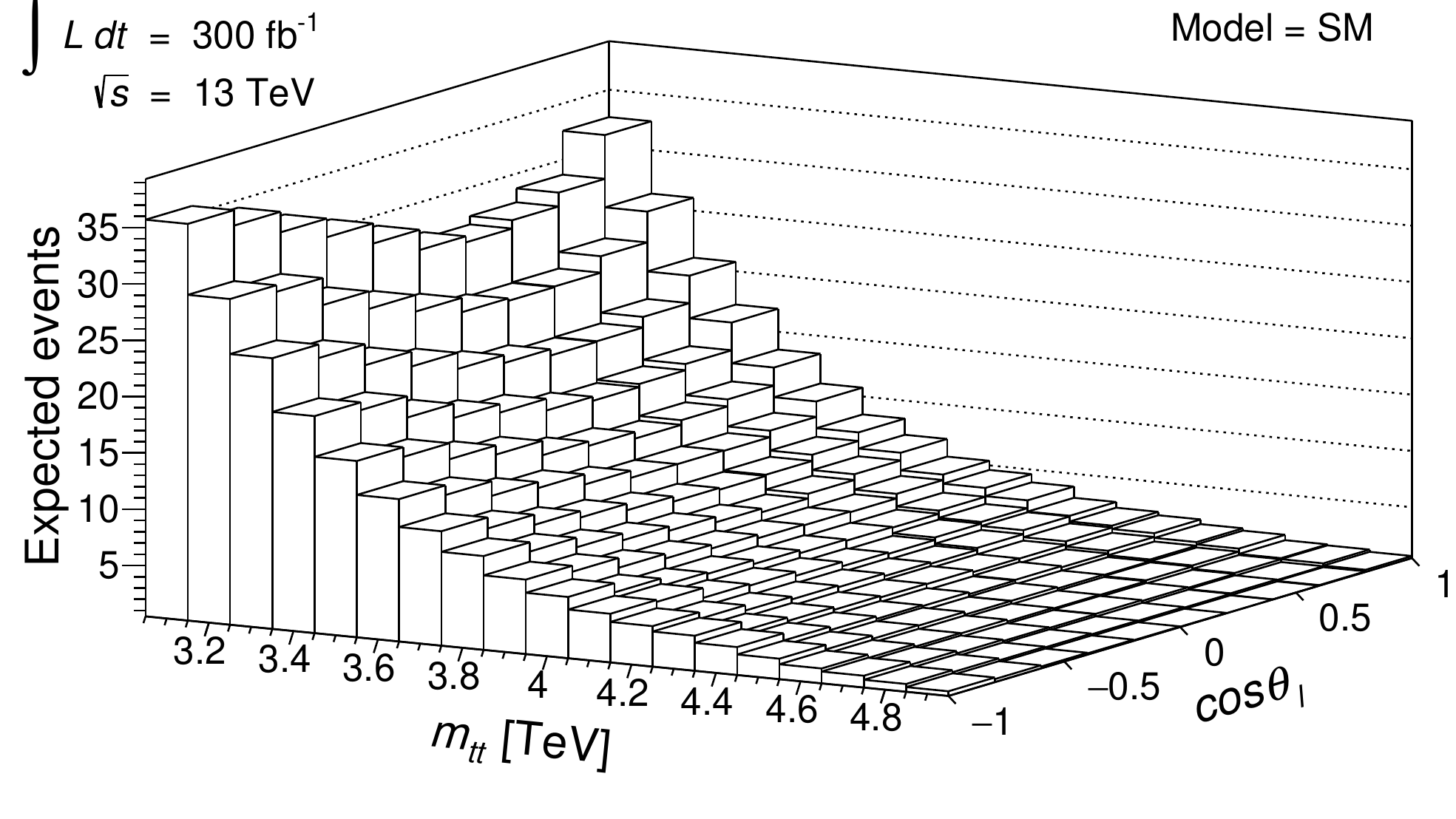}
    \caption{SM}
    \vspace{1cm}
  \end{subfigure}
  \begin{subfigure}{0.49\textwidth}
    \includegraphics[width=\textwidth]{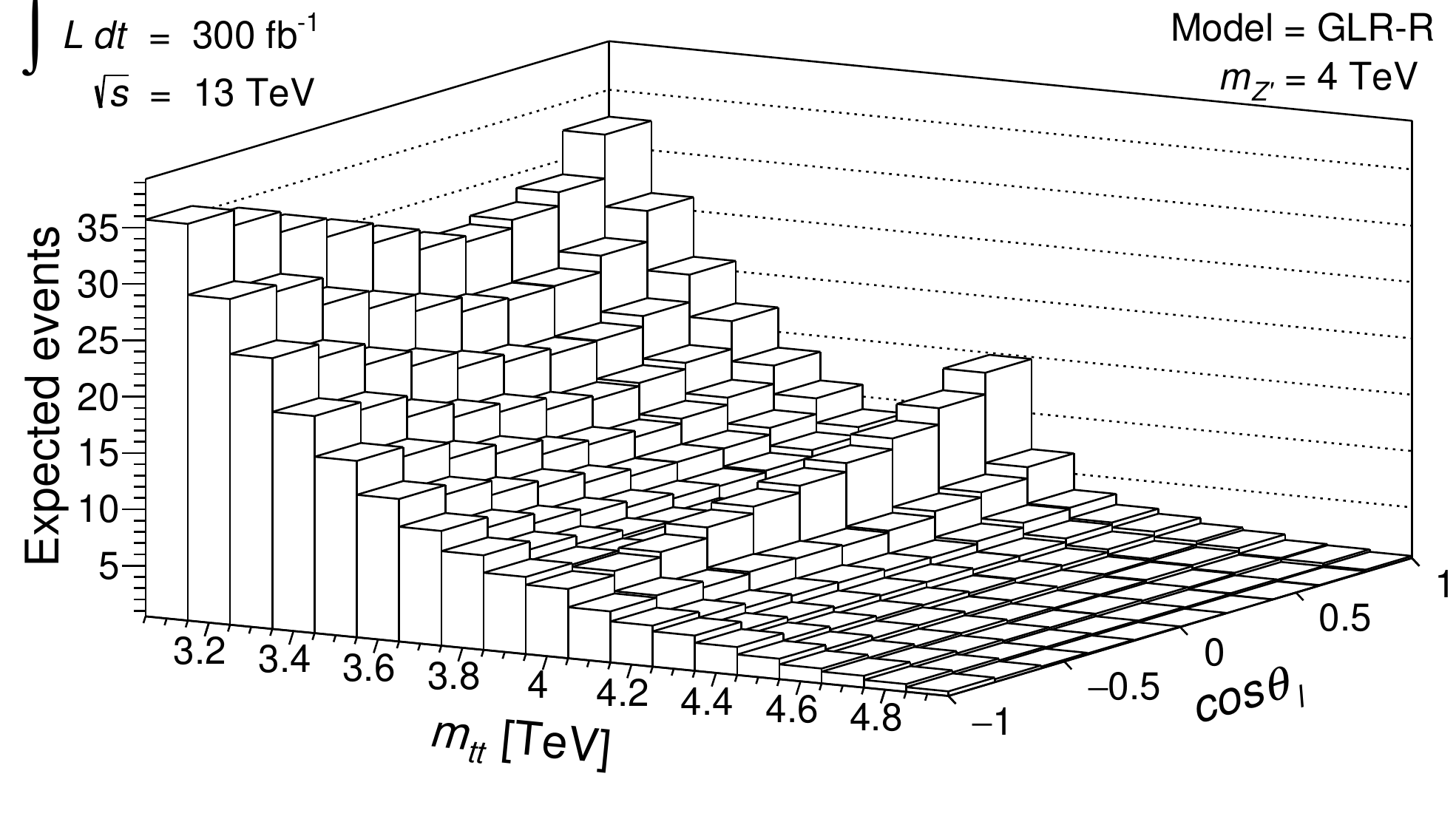}
    \caption{GLR-R}
    \vspace{1cm}
  \end{subfigure}
  \begin{subfigure}{0.49\textwidth}
    \includegraphics[width=\textwidth]{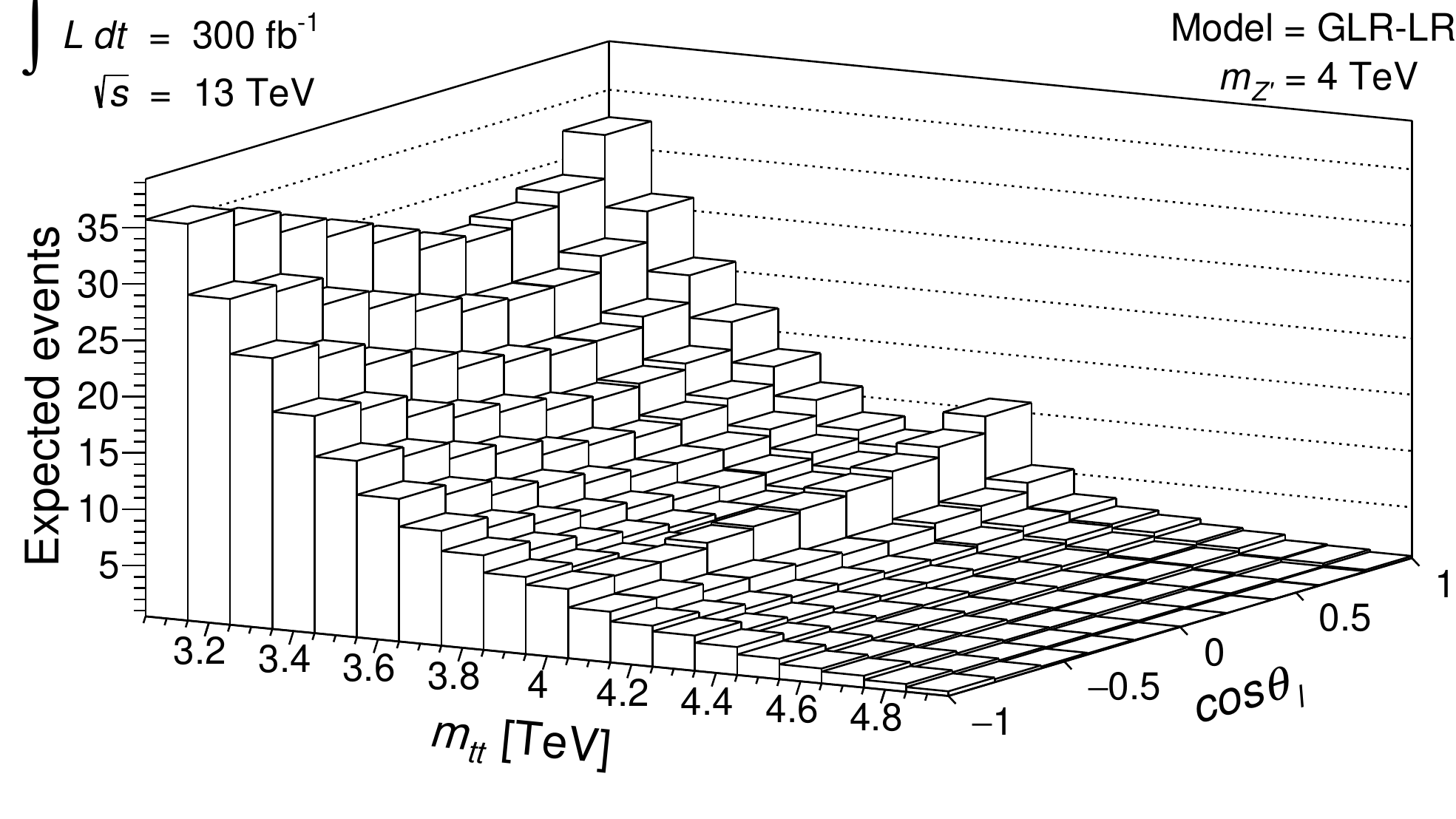}
    \caption{GLR-LR}
    \vspace{1cm}
  \end{subfigure}
  \begin{subfigure}{0.49\textwidth}
    \includegraphics[width=\textwidth]{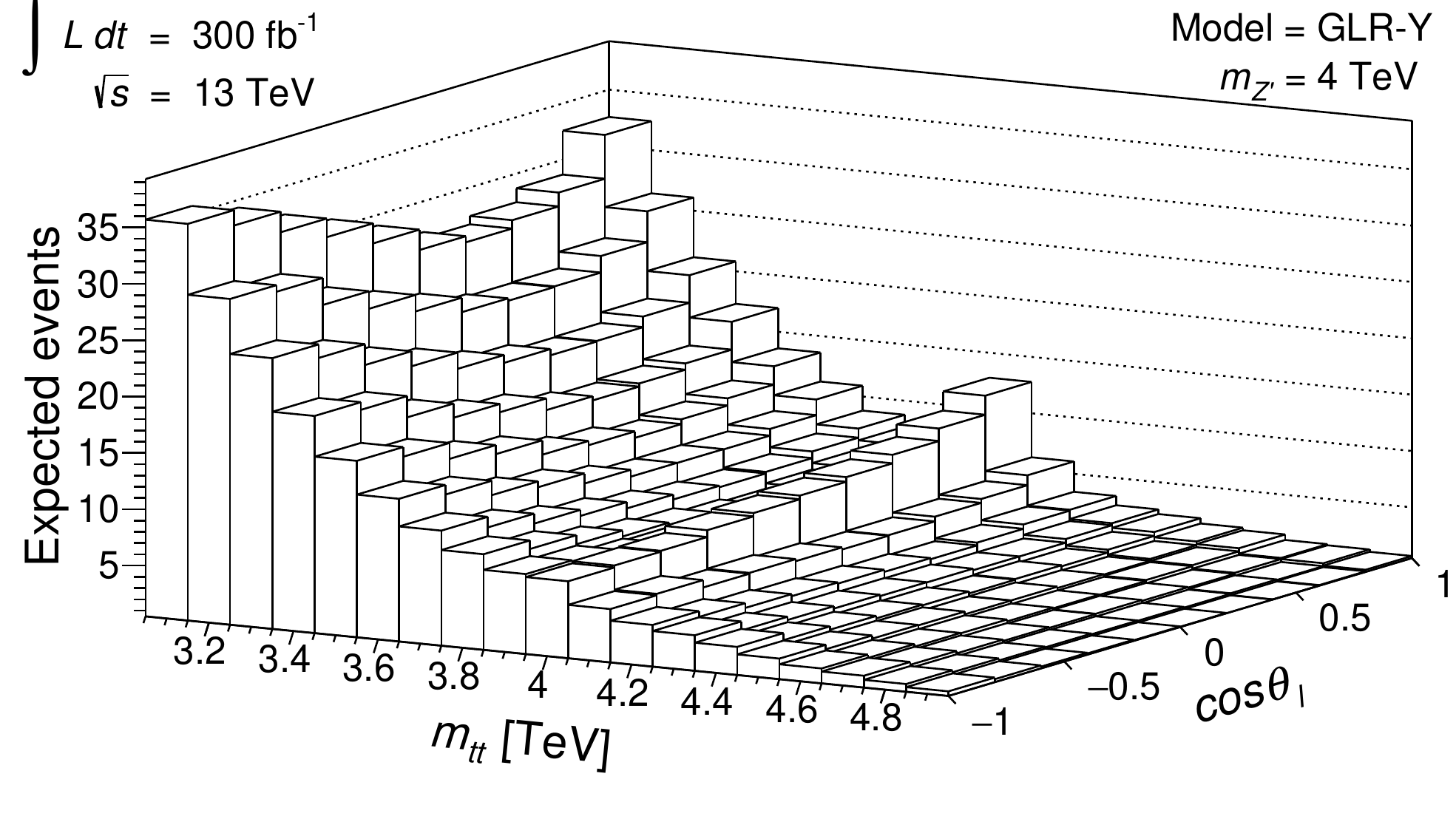}
    \caption{GLR-Y}
    \vspace{1cm}
  \end{subfigure}
  \begin{subfigure}{0.49\textwidth}
    \includegraphics[width=\textwidth]{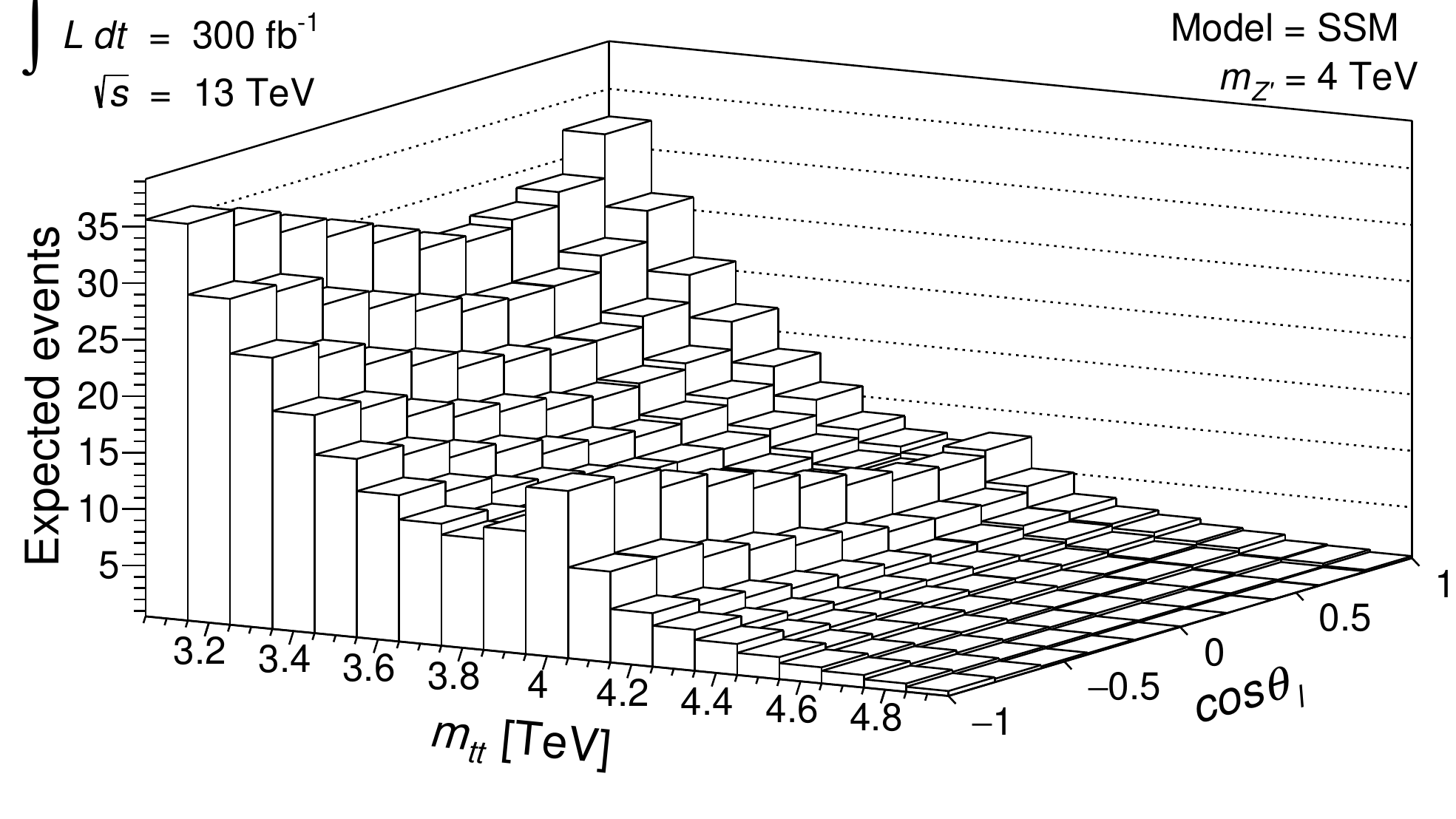}
    \caption{GSM-SM}
    \vspace{1cm}
  \end{subfigure}
  \begin{subfigure}{0.49\textwidth}
    \includegraphics[width=\textwidth]{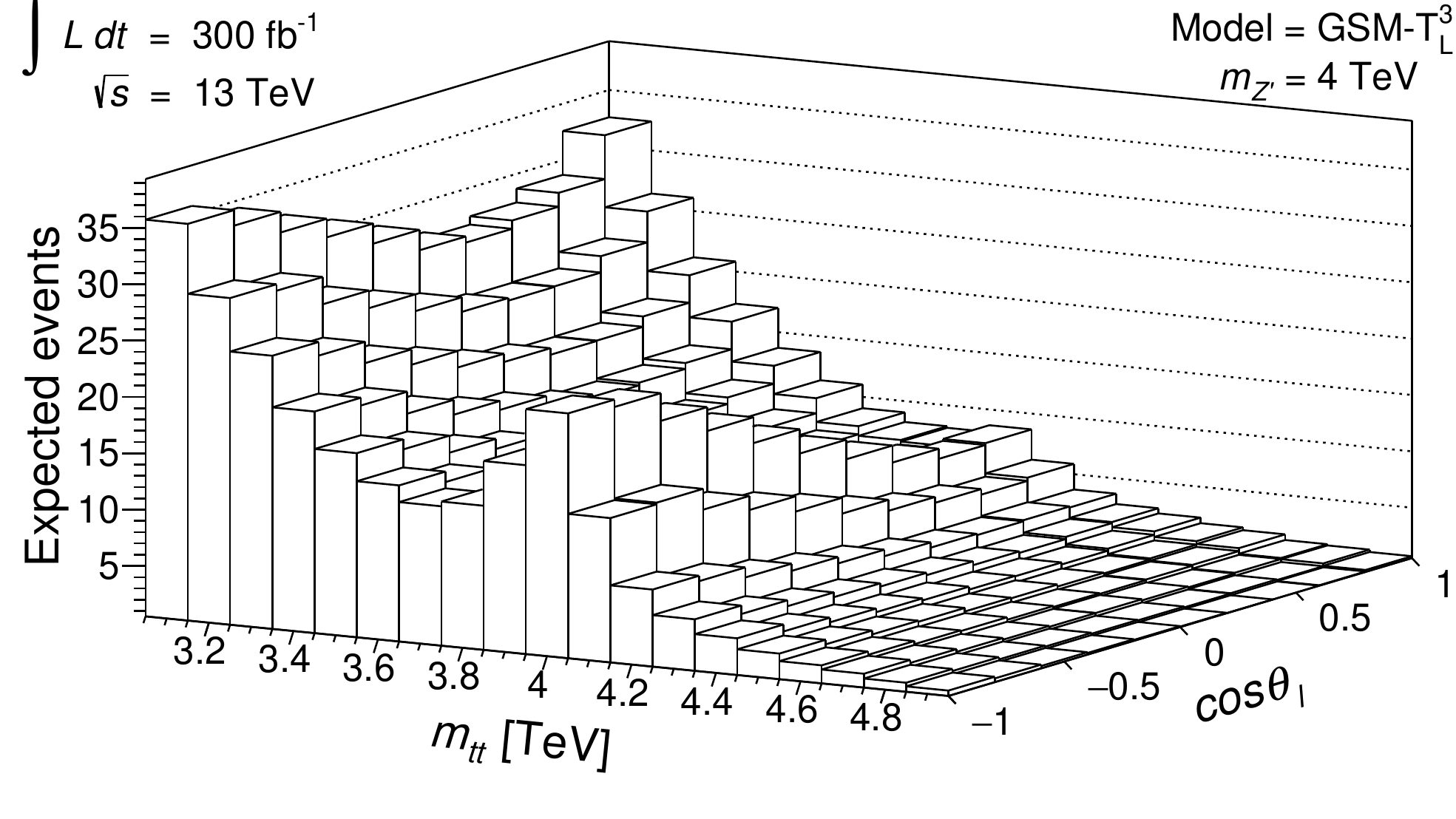}
    \caption{GSM-$T^3_L$}
    \vspace{1cm}
  \end{subfigure}
  \caption{Expected number of events binned in both $m_{tt}$ and $\cos\theta_{\ell}$ after toy reconstruction. These plots are used to derive the 2D significances in these variables, and access the combined significance of using the $m_{tt}$ distribution and $A_{L}$. Note the effect of the $Z'$ around $4$~TeV. The gradient of this slope is taken to extract $A_L$ (after normalisation).}
  \label{fig:mtt_costhetal}
\end{figure}

In the previous subsection, the capacity for asymmetries to diagnose a previously discovered $Z'$ candidate was emphasised. Now their potential to contribute to first detection of a $Z'$ particle in the $t\bar{t}$ channel (hence of discovery in certain models) is studied. To do this the generated events are binned in both $m_{tt}$ and the defining variable of the asymmetry: $\cos\theta^*$ for $A^t_{FB^*}$ and $\cos\theta_\ell$ for $A_L$. Ten bins are used across the range $-1 \leq \cos\theta \leq +1$. The resulting 2D histograms are presented in Figs.~\ref{fig:mtt_costhetastar} and \ref{fig:mtt_costhetal}, respectively.

The final results of the likelihood-based test, using asymptotic formulae, as applied to each model and tested against the SM are presented in Tab.~\ref{tab:significance}. The standard $m_{tt}$ based bump-hunt shows that the GLR and GSM models generally report a higher significance than those of the E$_6$ class. Of this class, the $\eta$ model shows the best potential for observation. Further support for an E$_6$ derived $Z'$ would be manifest in an accompanying negligible response in $A_L$ and $A^{t}_{FB^*}$.

The models with non-trivial asymmetries consistently show a potential for increased significance for the 2D histograms compared with using the cross section alone, illustrating their potential application in gathering evidence to herald the discovery of new physics. Additionally, in general, using $\cos\theta_{\ell}$ increases the significance more for GLR models than when using $\cos\theta^*$, while for GSM models the latter observable provides more sensitivity as a complimentary discovery observable. Combined with the potential for $A_L$ to distinguish between different classes of models, $\cos\theta_{\ell}$ represents the most interesting additional information when combined with the differential cross section.

\begin{table}
  \centering
  \begin{tabularx}{0.555\textwidth}{llrrr}
    \toprule
    Class & U$(1)'$ & \multicolumn{3}{c}{Significance ($Z$)} \\
    \cmidrule{3-5}
    & & $m_{tt}$ & $m_{tt}$ \& $\cos\theta^{*}$ & $m_{tt}$ \& $\cos\theta_{\ell}$ \\
    \midrule
    \multirow{4}{*}{GLR}   & $R$     &  $8.1$ &  $8.7$ &  $9.1$ \\
                           & $BL$    &  $1.4$ &  $1.4$ &  $1.4$ \\
                           & $LR$    &  $5.3$ &  $5.7$ &  $5.9$ \\
                           & $Y$     &  $7.1$ &  $7.6$ &  $7.7$ \\
    \midrule
    \multirow{3}{*}{GSM}   & $T^3_L$ & $16.9$ & $17.9$ & $17.5$ \\
                           & $SSM$   &  $9.9$ & $10.5$ & $10.2$ \\
                           & $Q$     & $30.2$ & $31.4$ & $30.3$ \\
    \midrule
    \multirow{6}{*}{E$_6$} & $\chi$  &  $1.1$ &  $1.1$ &  $1.1$  \\
                           & $\psi$  &  $3.7$ &  $3.9$ &  $3.7$ \\
                           & $\eta$  &  $6.6$ &  $6.9$ &  $6.6$ \\
                           & $S$     &  $0.1$ &  $0.1$ &  $0.1$ \\
                           & $I$     &  $0.0$ &  $0.0$ &  $0.0$ \\
                           & $N$     &  $1.5$ &  $1.6$ &  $1.5$ \\
    \bottomrule
  \end{tabularx}
  \caption{Expected significance, expressed as the Gaussian equivalent of the $p$-value. The E$_6$, U$(1)_{B-L}$ and U$(1)_{Q}$ models have negligible $A^{t}_{FB^{*}}$ and $A_{L}$, and thus no increase in significance when compared with the mass distribution alone.}
  \label{tab:significance}
\end{table}

\section{Conclusions}
\label{sec:conclusions}

In this paper, the scope of using the LHC to access semileptonic final states produced by $t\bar t$ pairs emerging from the decay of a heavy $Z'$ state above and beyond the SM background induced by both QCD and EW processes has been investigated. A variety of BSM scenarios embedding one such a state have been tested. The primary aim of this investigation was to extend earlier results produced limitedly to on-shell $t\bar t$ production which claimed that charge and spin asymmetry observables can be used to aid the diagnostic capabilities provided by the cross section in identifying the nature of a possible $Z'$ signal. This was done by adopting a $2\to 6$ calculation accounting for all possible topologies for both signal and (irreducible) backgrounds, including interference effects where appropriate.

Additionally, a statistical procedure was devised which enabling the combination of both differential and integrated significances from cross section and asymmetry signals in a two-dimensional analysis to show that, for models with non-negligible asymmetries, the significance of first discovery may be enhanced compared with using the cross section binned in one dimension. While the analysis was performed at the parton level, it included a reconstruction procedure of the top and antitop quark masses and momenta that closely mimics experimental conditions.

These results suggest that charge and spin asymmetry observables can have a strong impact in accessing and profiling $Z' \to t\bar{t}$ signals during Run 2 of the LHC, even aiding in first discovery. This is all the more important in view of the fact that several BSM scenarios, chiefly those assigning a composite nature to the recently discovered Higgs boson, embed one or more $Z'$ state which are strongly coupled to top quarks.

Therefore, the stage is set for a fully-fledged analysis eventually also to include parton-shower, fragmentation/hadronisation, heavy flavour decay and (true) detector effects, which will constitute the subject of a forthcoming publication. A core requirement, in pursuit of such an analysis, is the need to perform an appropriate boosted reconstruction that preserves high signal efficiency in the face of increasing momentum and maintains control over associated systematic uncertainties which, given the special phase space region, can suffer from specific limitations (e.g. control regions for validation and determination of scale uncertainties can be depleted of events and modelling uncertainties in generator and hadronisation can also play a role).

\section*{Acknowledgements}

We acknowledge the support of ERC-CoG Horizon 2020, NPTEV-TQP2020 grant no. 648723, European Union. DM is supported by the NExT institute and an ATLAS PhD Grant, awarded in February 2015. SM is supported in part by the NExT Institute and the STFC Consolidated Grant ST/L000296/1. FS is supported by the STFC Consolidated Grant ST/K001264/1. We would like to thank Ken Mimasu for all his prior work on $Z'$ phenomenology in $t\bar{t}$, as well as his input when creating the generation tools used for this analysis. Thanks are also due to Juri Fiaschi for helping us to validate our tools against his, in the case of Drell-Yan production. Additionally, we are very grateful to Glen Cowan for discussions on the statistical procedure, and Lorenzo Moneta for aiding with the implementation.

\bibliography{paper}{}
\bibliographystyle{paper}
\end{document}